\documentclass[12pt,a4paper]{article}

\usepackage{soul}
\usepackage{comment}
\usepackage{amsmath,amssymb}
\usepackage{verbatim}
\usepackage{graphicx}
\usepackage{mathrsfs,yfonts}
\usepackage{latexsym}

\usepackage{amsthm}
\usepackage{braket}
\usepackage{tikz}
\usepackage{subfigure}
\usepackage{aas_macros}
\usepackage[font=small]{caption}

\usepackage[nottoc]{tocbibind}

\usepackage[utf8]{inputenc}

\usepackage{amssymb}
\usepackage[hidelinks]{hyperref}
\usepackage{color}
\usepackage{cite}
\usepackage{footmisc}
\usepackage{geometry}
\usepackage{authblk}

\graphicspath{{./}{figures/}}

\usepackage{colortbl}
\definecolor{summersky}{cmyk}{0.71,0.33,0,0.14}
\definecolor{rp}{cmyk}{0.2, 1, 0.6, 0}

\newcommand\anote[1]{\textcolor{rp}{\bf [Ali:\,#1]}}

\newcommand{\orcid}[1]{\href{https://orcid.org/#1}{\textcolor[HTML]{A6CE39}{\aiOrcid}}}

\newcommand\ee{\end{equation}}
\newcommand\be{\begin{equation}}

\def\O{\mathcal{O}}
\title{Primordial black hole formation in matter domination}

\author[a,c]{E. Ebrahimian}
\author[b]{A. Abolhasani}
\author[a]{M. Mirbabayi}

\affil[a]{The Abdus Salam ICTP, Strada Costiera 11, I-34014 Trieste, Italy}
\affil[b]{Sharif University of Technology \\
Azadi Ave.\\
Tehran, Iran}
\affil[c]{IFPU, Institute for Fundamental Physics of the Universe, 34014 Trieste, Italy}
\date{}

\begin{document}

\maketitle

\begin{abstract}
  We study Primordial Black Holes (PBHs) formed by the collapse of rare primordial fluctuations during an early period of Matter Domination. The collapse threshold strongly depends on the shape of the peaks, decreasing as they become flatter and hence rarer. In the extreme limit of a top-hat perturbation, Harada, Kohri, Sasaki, Terada, and Yoo have argued that the growth of velocity dispersion prevents the formation of black holes unless the initial peak is larger than $\zeta_{\rm th} \sim \zeta_{\rm rms}^{2/5}$. Including the shape distribution of the peaks, we find that for a realistic cosmic abundance of PBHs, the effective threshold is larger, $\zeta_{\rm th} \sim \zeta_{\rm rms }^{1/10}$. And this model requires $\zeta_{\rm rms}\sim 10^{-1}$, which is much larger than the observed value at the CMB scales. Hence, PBH formation during Matter Domination is barely more efficient than Radiation Domination. We estimate the dimensionless spin parameter to be $a_{\rm rms} \sim \zeta_{\rm rms}^{7/4}\ll 1$, slightly larger than PBHs formed in Radiation Domination.

 
\end{abstract}

\section{Introduction}

The possibility of black hole formation due to the collapse of a large cosmological perturbation got the attention of physicists from the 1970s\cite{Zeldovich:1967lct,Carr:1975qj,Carr:1974nx,Hawking:1971ei}. Such black holes with non-stellar origin are called Primordial Black Holes (PBHs)\cite{Escriva:2022duf}. It is believed that these PBHs can account for some of the super-massive black holes \cite{1984MNRAS.206..801C,Bean:2002kx} or a part of the dark matter \cite{Chapline:1975ojl,Carr:2016drx,Carr:2020xqk}. Recent gravitational wave observations, \cite{LIGOScientific:2016dsl,KAGRA:2021vkt}, have also become another source of interest for PBHs since some of the gravitational wave events might involve black holes of primordial origin \cite{Carr:2023tpt}.

Several aspects of PBHs have been discussed in detail, including the minimum amplitude of primordial perturbation (denoted by $\zeta$) required to form a black hole \cite{Shibata:1999zs,Musco:2012au,Polnarev:2006aa,Harada:2015yda,Escriva:2020tak}, abundance and mass function \cite{Carr:2017jsz,Kuhnel:2017pwq,Germani:2018jgr,Young:2019yug}, spin \cite{Mirbabayi:2019uph} and other properties. Most of these studies are carried out for a general perfect fluid with nonzero pressure, in other words, for $w\neq 0$, where $w=p/\rho$. Generally, lower values of $w$ lead to lower thresholds for PBH formation. The reason for this is the role of the pressure gradient in defending against gravity.

However, these analytical approaches to determine the black hole formation threshold lead to $\zeta_{\rm th}=0$ for $w=0$ \cite{Carr:1975qj,Harada:2013epa}. If correct, this would lead to a very different-looking universe than the one we live in by implying that perturbations during the late matter-dominated era would collapse into black holes rather than Virialized dark matter halos. In the context of PBH formation, it is particularly relevant to address this question in the context of inflationary models that end with an early matter-dominated era before reheating. See \cite{Green:1997pr,Easther} for some examples.

In the case of halo formation during late matter domination, velocity dispersion plays a crucial role in preventing small perturbations from collapsing into black holes, instead transforming them into Virialized structures. This argument has recently been used in \cite{Harada:2022xjp} to estimate the PBH threshold. They consider a {\em top-hat} overdensity, superposed by typical inhomogeneities characterized by $\zeta_{\rm rms}\ll 1$ at an early time. By studying the growth of perturbations in a dust-dominated universe, they find a black hole threshold $\zeta_{\rm th}\sim \zeta_{\rm rms}^{2/5}$. For small $\zeta_{\rm rms}$ this threshold is smaller than $1$, which supports the idea that the growth of overdensities at subhorizon scales enhances PBH formation. On the other hand, $\zeta_{\rm rms}^{2/5} \gg \zeta_{\rm rms}$, which guarantees that only an exponentially small fraction $\beta$ of peaks collapse into black holes while most peaks collapse into halos. More explicitly $\log\beta \sim  -\zeta_{\rm rms }^{-6/5}$.

Even with this more conservative threshold, there is still an overestimation of primordial black hole (PBH) formation. The top-hat profile is unique because, in the absence of fluctuations, it will inevitably collapse into a black hole. However, a top-hat configuration is highly unlikely. A typical overdensity, even if perfectly spherically symmetric like a Gaussian, will experience shell-crossing and Virialization in a dust-dominated universe unless its initial amplitude is $\O(1)$.

We will see that as the profile becomes flatter and less typical, the threshold for collapse gradually decreases. The flatness of the profile can be quantified by the smallness of the spatial derivatives at the peak. There is a competition between the enhancement due to the lower PBH-formation threshold and the suppression due to the atypicality of these derivatives, which for a given $\zeta_{\rm rms}$ determines the most likely PBH formation history. We will find that for a cosmologically relevant value of $\beta$, we need $\zeta_{\rm rms}\sim 0.1$ for which most PBHs are formed from peaks with significantly small second derivative, but typical higher order derivatives. They are significantly more common than a top-hat but significantly less flat, resulting in an effective threshold $\zeta_{\rm th} \sim \zeta_{\rm rms}^{1/10}$, or
\be
\log\beta \sim -\zeta_{\rm rms} ^{-1.8}.
\ee
The rest of this paper provides the argument leading to the derivation of the above result and addresses several apparent puzzles and questions in this context. In particular, in Section \ref{sec:RV}, we review the standard formalism of PBH formation from the collapse of {\em spherically symmetric} peaks. In Section \ref{sec:wsol}, we point out the apparent distinction between dust and $w \to 0$ limit of a fluid: the fact that black hole formation is very sensitive to the profile shape in the dust scenario whereas in a fluid with constant $w$, most spherically symmetric peaks collapse into black holes when $w\to 0$. In Section \ref{sec:Pertb}, we analyze the influence of density perturbations on PBH formation and suggest that minor deviations from spherical symmetry can reconcile the dust and fluid models. Section \ref{sec:NumAnl} presents numerical evidence demonstrating how shell-crossing and velocity dispersion stop the collapse and lead to Virialization. Based on this framework, Section \ref{sec:Abund} estimates the abundance and clustering of such PBHs, while Section \ref{sec:Spin} estimates their spin. Contrary to earlier claims \cite{Harada:2017fjm}, we find that the spin is typically small, essentially because the total angular momentum, which requires a net rotation, plays a subdominant role compared to velocity dispersion in PBH formation. We conclude with a summary and discussion in Section \ref{sec:Con}.

{\em Notation:} Here we set $c=1$, and when we write $X_k$ for some field, we mean $X_k=\sqrt{k^3 P(k)}$ where $P(k)$ is the Fourier transformation of the two-point function of $X(\boldsymbol{r})$. For a scale-invariant spectrum, we would have $X_k=\mathrm{const.}$

\section{Review of PBH formation formalism}\label{sec:RV}
PBHs form from the collapse of rare perturbations. Typically, these rare fluctuations are nearly spherically symmetric. For instance, in a Gaussian random field $\zeta$, the largest and smallest eigenvalues of the Hessian matrix at a rare peak satisfy\cite{BBKS}
\begin{equation}
\frac{\lambda_{\rm max}-\lambda_{\rm min}}{\lambda_{\rm max}+\lambda_{\rm min}} \sim \frac{\zeta_{\rm rms}}{\zeta_{\rm peak}},
\end{equation}
which can be interpreted as a large spherically symmetric perturbation of height $\zeta_{\rm peak}\gg \zeta_{\rm rms}$ superposed by typical perturbations with the amplitude of $\zeta_{\rm rms}$. In this work, we assume negligible primordial non-Gaussianity and adopt this picture, with $\zeta$ being the adiabatic curvature perturbations (see Appendix \ref{sec:sphr} for more discussion).

Hence, we will start by reviewing the evolution of a perfectly spherical perturbation in a cosmology dominated by a perfect fluid with equation of state $p = w \rho$. Ultimately, we will be interested in the subhorizon evolution in the limit $w\to 0$,  where an analytic solution for spherical collapse exists. The following discussion allows us to match the parameters of that solution with the adiabatic initial conditions.

In general, any spherical perturbation can be described by the following metric:
\begin{equation}\label{eq:sphericmetric}
ds^2=-A^2(r,t) dt^2+B^2(r,t)dr^2+R^2(r,t) d\Omega^2.
\end{equation}
$A$, $B$, and $R$ are functions of radial coordinate $r$ and time $t$. We work in the comoving gauge, where $R(r,t)$ represents the areal radius of the matter distribution at time $t$. It is useful to introduce the following dynamical variables:
\begin{equation}
    U:=A^{-1}\dot{R},\quad M:=\int_0^r 4\pi \rho\,R^2 R' dr ,\quad \Gamma^2:=\frac{R'^2}{B^2},
    \label{eq:MS-defs}
\end{equation}
where $U$ represents the radial velocity of matter, $M$ is the Misner-Sharp mass, and $\Gamma$ corresponds to a generalized Lorentz factor that combines special relativistic effects with gravitational redshift. Here overdot is time derivative, and prime denotes the derivative with respect to $r$. Using these variables and applying the Einstein equations to the metric \eqref{eq:sphericmetric} for a perfect fluid with equation of state $p = w\rho$, we obtain the Misner-Sharp equations \cite{PhysRev.136.B571}:
\begin{align}
&\Gamma^2=1+U^2-\frac{2GM}{R}\\
& \dot{M}=-4\pi R^2\rho w\, U A\\
& A'=-A\frac{\rho'w}{\rho(1+w)}\\
&\dot{\rho}=-A\rho(1+w)\Big(\frac{2U}{R}+\frac{U'}{R'}\Big)\\
&\dot{U}=-A\left(4\pi{G} R \rho w+\frac{GM}{R^2}+\frac{w\rho'}{\rho(1+w)}\frac{\Gamma^2}{R'}\right).
\end{align}
To determine the evolution of the metric coefficients, one must solve the coupled system of Misner-Sharp equations. Although these equations generally require numerical treatment, analytical solutions exist in certain limiting cases. A particularly important limit occurs for super-horizon perturbations, where spatial gradients become negligible. {By neglecting the sub-leading terms in the $t\to 0$ limit} the metric can be expressed in either of two equivalent forms:
\begin{equation}
\label{eq:metric-Kr}
ds^2 = -dt^2 + \frac{a^2(t) dr^2}{1-K(r)r^2} + a^2(t) r^2 d\Omega^2,
\end{equation}
or alternatively:
\begin{equation}
ds^2 = -dt^2 + a^2(t) e^{2\zeta(\tilde{r})} (d\tilde{r}^2 + \tilde{r}^2 d\Omega^2),
\end{equation}
where $a(t)$ represents the background FLRW scale factor, $K(r)$ and $\zeta(\tilde{r})$ characterizes the curvature perturbation, subjected to the boundary condition at infinity, that the universe be flat, i.e. $\displaystyle{\lim_{r\to \infty}} K(r)r^2=0$, or equivalently $\displaystyle{\lim_{\tilde{r}\to \infty}} \zeta(\hat{r})=0$. The two $r$ and $\tilde{r}$ coordinates are related by the following equations:
\begin{equation}
r=\tilde{r}e^{\zeta(\tilde{r})},\quad \tilde{r}=r\exp\int_{\infty}^{r}\frac{d\bar{r}}{\bar{r}}\Big(\frac{1}{\sqrt{1-K(\bar{r})\bar{r}^2}}-1\Big).
\end{equation}
$\zeta$ and $K$ are non-linearly related to each other:
\begin{equation}
    (1+\tilde{r}\zeta')^2=1-K(r)r^2.
\end{equation}
 However, for small perturbations we have\emph{ $\tilde{r}\approx r (1+\mathcal{O}(\zeta))$, so to the first order we have}:
\begin{equation}\label{eq:zetaKrelation}
    \zeta(\tilde{r})\approx \frac{1}{2}\int_{\tilde{r}}^{\infty}dr rK(r).
\end{equation}
 $\zeta(\tilde{r})$ and $K(r)$ are adiabatic metric perturbations that serve as the initial conditions for the Misner-Sharp equations \cite{Polnarev:2006aa}. To connect the super-horizon limit with the full Misner-Sharp system, one could employ the gradient expansion method, expanding the equations of motion linearly in terms of a small parameter $\epsilon$ that characterizes spatial derivatives (see \cite{Harada:2015yda}). If $r_m$ represents the spatial size of the perturbation, then:
\begin{equation}
    \epsilon=\frac{H(t)^{-1}}{a(t)r_m}
\end{equation}
where $H(t)$ is the Hubble parameter. The condition $\epsilon\ll 1 $ corresponds to perturbations outside the horizon. By expanding physical quantities in terms of $\epsilon$ and solving the Misner-Sharp equations, we obtain solutions that are valid for small $\epsilon$: \footnote{The expansion starts from $\epsilon^2$ because they must connect to growing modes after inflation, see \cite{Harada:2015yda}. }
\begin{align}
&R(t,r)=a(t)r(1+\epsilon^2 \tilde{R}),& &\tilde{R}=\frac{\tilde{U}}{1+3w}-\frac{w\tilde{\rho}}{(1+3w)(1+w)},\\
&U(t,r)=H(t)R(1+\epsilon^2\tilde{U}),& &\tilde{U}=\frac{-1}{5+3w}K(r) r_m^2,\\
&\rho(t,r)=\rho_b(t)(1+\epsilon^2\tilde{\rho}),& &\tilde{\rho}=\frac{1+w}{5+3w}\frac{r_m^2}{r^2}(K(r)r^3)',\\ \label{eq:Mepsilon}
&M=\frac{4\pi}{3}\rho_b(t)R^3(1+\epsilon^2\tilde{M}),& &\tilde{M}=-3(1+w)\tilde{U}.
\end{align}
Here $\rho_b(t)$ is the background density. This shows that $K(r)$ will determine the initial condition for the equations. The studies that determine the PBH formation threshold use these equations to set the initial condition of the Misner-Sharp equations\cite{Polnarev:2006aa}. They assume a sufficiently small $\epsilon$ at some initial time, meaning that the perturbation is well outside the horizon at that time, and then numerically solve the nonlinear equations \cite{Shibata:1999zs,Escriva:2019nsa}. Black hole formation is identified by the appearance of an apparent horizon, which occurs when the areal radius satisfies:{ $R(r,t) = 2GM(r,t)$ for a radius.} 

It is customary to express the black hole formation threshold in terms of {\it compaction function} defined as \cite{Shibata:1999zs}:
\begin{equation}
\mathcal{C}(r,t)=2\frac{G(M-M_b)}{R};\quad M_b=\frac{4\pi}{3}\rho_b R^3.
\end{equation}
On super-horizon scales, the compaction function becomes constant and proportional to the  curvature profile:
\begin{equation}
\mathcal{C}(r,t=0) =  \frac{3(1+w)}{5+3w} K(r) r^2.
\end{equation}
The maximum of the above function is defined as the amplitude of the perturbation $\delta_m$. The exact definition of $r_m$ is where the compaction function reaches its maximum: $\partial_r\mathcal{C}(r,0)|_{r=r_m}=0$, and the amplitude of the perturbation is:
\begin{equation}\label{eq:deltamdef}
    \delta_m=\mathcal{C}(r_m,0).
\end{equation}
{For perturbatively small peaks, $\delta_m\ll 1$, we have $\zeta_{\rm peak}\sim \delta_m\sim K(0)r_m^2$.} For example, for a Gaussian profile of $K(r)$, we have $K(r)=\frac{5\delta_m}{3r_m^2} e^{1-(r/r_m)^2}$. Using Eq.\eqref{eq:zetaKrelation}, which is valid in perturbative regime, $\zeta(\tilde{r})\approx \frac{5\delta_m}{12}e^{1-(\tilde{r}/r_m)^2}$ and $\zeta_{\rm peak}=\zeta(0) \approx 1.13 \delta_m$. 

The threshold of PBH formation, $\delta_{\rm th}$, is defined as the minimum amplitude of a density perturbation required for gravitational collapse to result in black hole formation. Previous studies, both numerical and analytical, have shown that $\delta_{\rm th}\to 0$ when $w\to 0$ \cite{Escriva:2020tak,Harada:2013epa}.  We will argue that this result is a consequence of two assumptions: spherical symmetry and perfect fluid approximation. {While the simulations carried out with those assumptions correctly reflect the PBH formation threshold within those assumptions, for $w\ll 1$ the applicability of those assumptions, and thus the results of those simulations, to finding the threshold of PBH formation depends on the perturbation profile, the microscopic matter model, and, more importantly, the amplitude of the power spectrum. This is because our analysis suggests that, for sufficiently small $w$, nonspherical perturbations and the associated velocity dispersion or effective anisotropic stresses become dynamically important. We will argue that in the phenomenologically interesting PBH scenarios, where the amplitude of the power spectrum is fixed by the desired PBH abundance, non-sphericities are large and cannot be neglected for $w\ll 1$.}

\section{Spherical collapse in dust, and the $w\to0$ limit}\label{sec:wsol}

When considering fluids with small $w$, the reasoning behind the linear scaling of $\delta_{\rm th}\propto w$ is quite straightforward. A small overdensity, with initial value $\delta_m$, enters the horizon and grows until it decouples from the background expansion and starts to collapse at the ``turnaround'' time. At this point $\delta_m$ is approximately the same as the gravitational potential. This is because the gravitational potential is conserved during the whole course of linear growth. Therefore, the ratio between the Jeans length, $\lambda_J$, and the physical size of the perturbation, $R_m$, at this time is roughly
\begin{equation}
\left.    \frac{\lambda_J}{R_m}\right|_{\rm turnaround}\sim \sqrt{\frac{w}{\delta_m}}. 
\end{equation}
When $w > \delta_m$ pressure gradients win over gravitational forces and prevent the collapse. In the opposite regime then Jeans instability occurs: the Jeans length will shrink $\propto R_m^{3/2}$ thus it always remains smaller than $R_{m}$ and can't prevent collapse after it starts.\footnote{Note that this is only the case for constant $w$, for example in baryonic matter $w$ itself will grow and at some point the Jeans length can be comparable with the structure size after collapse.} This explains the linear relation between the threshold and $w$:
\begin{equation}\label{eq:linthreshol}
    \delta_{th}\approx Cw;\quad w\ll1,\quad 1\lesssim C.
\end{equation}
$C$ is a constant only to ensure that $\delta_m$ is large enough to start the collapse at the turnaround time. This is in agreement with the analytical estimate of \cite{Harada:2013epa} where $C\approx 3\pi^2/5$.

We will now study the dust-filled Universe, highlighting the impact of profile shape and shell crossing. The equations for dust are obtained by setting $w=0$. Then the Misner-Sharp equations have an analytic solution for early times. For the metric given by \eqref{eq:metric-Kr}, $\Gamma$ is a time constant, and $\Gamma^2 = 1 - K(r)r^2$. In this context, the energy constraint component of the Einstein equation — equivalent to the first Friedmann equation in the FRW geometry — reads as 
\begin{align}
\dot{R}^2(r,t)-\dfrac{2 G M(r)}{R(r,t)}=-K(r)r^2.
\end{align}
In the equation above, we utilized the fact that in a matter-dominated universe, with $w=0$, it follows that $\dot{M}(r,t)=0$. As a result, $M$ must be a function of $r$ only, which can be fixed by taking $t\to 0$ leading to $M\to M_b=\frac43\pi r^3 \rho_b a^3$. {This can be shown if we use Eq.\eqref{eq:Mepsilon} which leads to $M=\frac{4\pi}{3}r^3\rho_b a^3 (1+\epsilon(3\tilde{R}+\tilde{M}))$, and that when $w=0$ then $\tilde{M}=-3\tilde{U},\tilde{R}=\tilde{U}$, so the first order term in $\epsilon$ will cancel out}. Then one can find a parametric solution for $R(r,t)$ which is similar to a closed matter-dominated universe:
\begin{align}\label{eq:sphsol1}
 R(r,t)=\frac{GM(r)}{K(r)\,r^2}(1-\cos\theta)~~~~\implies&~~~  R(r,\theta)=\frac{H_0^2}{2K(r)}\,r(1-\cos\theta)\\
t(r,\theta)=\frac{GM(r)}{K^{\frac{3}{2}}(r)\, r^3}(\theta-\sin\theta)~~~\implies& ~~~~  t(r,\theta)=\frac{H_0^2}{2K^{\frac{3}{2}}(r)}(\theta-\sin\theta),\label{eq:sphsolt}
\end{align}
where $H_0$ is the expansion rate when $a(t_0) =1$ { and $\theta$ is called the implicit time to parameterize the solution}. {These are actually Lemaitre-Tolman-Bondi (LTB) solutions derived from a known curvature profiles.} The constancy of $M(r)$ also allows for determining the density at any given $r$ and $t$ using only the mass conservation law: $\rho(r,t)= M'(r)/(4\pi R^2R')\equiv\rho_b a^3 r^2/R^2R'$. {This leads to the following exact expression for the density profile:
\be\label{eq:rhoofr}
\rho(r,t)=\frac{8K(r)^3\rho_b a^3/H_0^6}{(1-\cos\theta)\Big((1-\cos\theta)^2(1-\frac{K'r}{K})+\frac{3K'r}{2K} \sin\theta(\theta-\sin\theta)\Big)}.
\ee}

The solution Eq.\eqref{eq:sphsolt} suggests that as the value of $K $ increases, the time required for collapse decreases. Consequently, for a monotonically decreasing profile, the inner layers tend to shrink more rapidly than the outer layers. This results in the profile transforming into a cuspy shape over time. We will return to this point in Section \ref{subsec:GB}.

This solution describes the evolution of an initial superhorizon perturbation $K(r)$. Initially, it expands with the background FLRW flow, but when the perturbation is an overdensity, various shells reach a maximum expansion point and turn around at $\theta = \pi$, shrinking afterward toward the origin as $\theta\to 2\pi$. For a monotonically decreasing curvature profile, at the time $t_c$ the density at the center diverges, where $t_c$ is:
\be\label{eq:t1equation}
t_c=\frac{\pi H_0^2}{K^{\frac{3}{2}}(0)}.
\ee
{One can see this by looking at Eq.\eqref{eq:rhoofr} and setting $r=0$, at $\theta=2\pi$ the density will diverge. Eq.\eqref{eq:sphsolt} shows that $\theta=2\pi$ at $r=0$ is equivalent to $t=t_c$. It is also possible that in Eq.\eqref{eq:rhoofr} the other more complicated term in the denominator, which involves $K'r$, vanishes before $t_c$ at a finite non-zero $r$. While this could happen for a general profiles, for a monotonically decreasing profile, i.e. $K'<0$, all the terms become positive and non-zero. The divergence of density at the center means that we cannot continue the evolution unless we somehow soften the divergence. {Here we must state that this is a breakdown of the LTB solution itself: the shell-focusing singularity $R(0,t_c)=0$, $\rho(0,t_c)\to\infty$ is the first obstruction to the classical evolution, and it is reached before any softening or coarse-graining is invoked.} Of course, in any realistic underlying model this singularity will be softened, for instance because the fluid consists of particles with finite mass, or, more importantly, there are anisotropies\footnote{{The term "anisotropies" in this manuscript, if used alone, always means profile anisotropies. This can develop into anisotropic dispersion velocities, but if we want to point out the dispersion velocity or other types of anisotropies, we will clearly state that.}}. {However note that such a softening, and the subsequent passage of the central shells through the origin, which will be discussed below, is a physically motivated \emph{continuation} of the LTB evolution for an effectively collisionless matter model.}}

{ Before turning to this continuation, it is important to check whether an apparent horizon has already formed during the classical LTB evolution, prior to $t_c$. Using Eq.~\eqref{eq:sphsol1}, one can show that:
\begin{equation}
    \frac{2GM}{R}=\frac{2K(r)r^2}{1-\cos\theta}.
\end{equation}
On the other hand from the definition of $\delta_m$ and the compaction function, one can show that $K(r)r^2\leq 5\delta_m/3$ so in general, and as long as LTB solutions are valid, we have: $\frac{2GM}{R}\leq \frac{10\delta_m}{3(1-\cos\theta)}$. This shows that if the condition $(1-\cos\theta)=0.3\delta_m$ is never satisfied, then no apparent horizon can form as long as LTB solutions are valid. Since we are working in the $\delta_m\ll 1$ regime, this could only happen if for some values of $r,t$ we have $\theta\approx 2\pi-\sqrt{0.6\delta_m}$, but plunging this into Eq.\eqref{eq:sphsolt} will result in $t=t_c (K(0)/K(r))^{3/2}[1-(0.6\delta_m)^{3/2}/12\pi]$, which, along with our previous assumption about the behavior of $K(r)$ with $r$, means that we should look for the formation of an apparent horizon only around the center and close to $t_c$. So we only need to look for an apparent horizon around the center at $t_c$.}

To this end, we must specify the functional form of $K(r)$. For $ r \ll r_m $, we can Taylor expand $ K(r) $ to $ K(0)(1-\sum_{n=1}^{\infty} A_{2n} (r/r_m)^{2n}) $, where $ A_{2n} $ are constants that depend on the shape of the profile. For a typical peak $A_2=\mathcal{O}(1)$ and we can ignore higher order terms when $r\ll r_m$ (see Appendix \ref{sec:sphr}). But it is instructive to see the dependence on $n$ and $A_{2n}$ for a simplistic monochromatic model like $K(r)=K(0)(1-A_{2n} (r/r_m)^{2n})$. At $t=t_c$, the central density diverges, but black hole formation requires evaluating the apparent horizon condition, {$R=2GM$}:
\begin{equation}\label{eq:TH1}
    \frac{2GM}{R}\Big|_{t_c}=\frac{2K(r)r^2}{1-\cos\theta}\sim \frac{\delta_m}{A_{2n}^{2/3}}\Big(\frac{r}{r_m}\Big)^{(6-4n)/3},
\end{equation}
where we used $K(0) \sim \delta_m/r_m^2$, and $2\pi - \theta(r,t_c) \sim A_{2n}^{1/3} (r/r_m)^{2n/3}$. {The latter is obtained by solving Eq.\eqref{eq:sphsolt} for $t=t_c$ and $r\ll r_m$. We have also neglected $\mathcal{O}(1)$ coefficients because we are only interested in the parametric dependence of $GM/R$ on the profile properties.} {Now we have two distinct cases. For a generic, $n=1$ profile, Eq.~\eqref{eq:TH1} shows that $2GM/R\to0$ as $r\to0$ at $t_c$, so that no trapped sphere exists arbitrarily close to the center; combined with the fact that we only expect to have an apparent horizon around the center at $t_c$, this means that to have a black hole in this case we must have $A_2^{2/3} \ll \delta_m$. For $n\ge2$, on the other hand, Eq.~\eqref{eq:TH1} implies that for any value of $\delta_m$ and $A_{2n}$ there is a finite $r$ which is close enough to $r=0$ that satisfies the apparent horizon condition. This $n\ge2$ case therefore corresponds to black hole formation \emph{during} the classical LTB evolution, before the shell-focusing singularity and independently of any assumption about its continuation.} All in all, unless the curvature profile is sufficiently flat, i.e. either $A_2^{2/3} \ll \delta_m$ or $n \geq 2$, an apparent horizon will not form by $t_c$.

{Can a black hole form afterwards (assuming the evolution is continued beyond $t_c$ after softening the divergence)? In Section \ref{sec:SimShell}, we demonstrate that the numerical simulations do not support this idea. Heuristically, after softening the density divergence at the origin, if the matter interacts only gravitationally, which we take as the definition of the dust model, the central layers that were moving inward before $t_c$ pass through the origin and move outward after $t_c$ and start crossing the in-going outer layers. This is an example of the ``shell-crossing'' phenomenon, though one that starts from the origin after softening the divergence.\footnote{{In the literature, ``shell-crossing'' is often used to describe two layers with different initial radii overtake one another during gravitational collapse. In our case, this phenomenon starts from the origin. After ``shell-crossing'', the perfect fluid description breaks down since there are two matter streams with different velocities at the same position in space.}} Afterwards, the out-flowing shells tend to reduce the density of the central region, and the infalling shells cannot compensate for this decrease. Another way of stating this is with the help of anisotropic pressure. After $t_c$, a nonzero radial velocity dispersion, $\sigma_r$, will arise and this significantly reduce the mean radial velocity, $\braket{v_r}$, around the center. {Since we have $\frac{\partial}{\partial t}\frac{GM}{R}=-4\pi G\rho R\braket{v_r}$, where $\partial/\partial t$ is time derivative at fixed areal radius, the reduction in the absolute value of $\braket{v_r}$ slows the growth of $GM/R$.} }

Hence, Eq.\eqref{eq:TH1} suggests that for a {\em typical} peak, namely one with $n=1$ and {$A_2\sim 1$,}  black hole formation threshold is {$\delta_{m}\sim 1$}. For a parametrically smaller $\delta_m$, spherical shell-crossing will prevent black hole formation unless the profile is sufficiently flat, i.e., $A_2\ll 1$. Of course, now one has to be more careful about (a) competition between different $r^{2n}$ terms in the Taylor expansion of $K(r)$, and (b) deviation from spherical symmetry due to the RMS fluctuations. We will discuss the role of the RMS fluctuation in the next section and find the threshold of PBH formation in the presence of perturbations.

{To summarize, the key difference between $w=0$ and $w\to 0^+$ that underlies the discrepancy in the black hole formation thresholds, $\delta_{\rm th} \sim \O(1)$ vs. $\O(w)$, is the possibility of shell-crossing, or in other words, the existence of anisotropic pressure. {In a perfect fluid scenario with $w\neq 0$, the pressure gradients always prevent shell-crossing and keep the fluid as a single stream, so the central density does not diverge before the formation of an apparent horizon. In other words, pressure gradients prevent the emergence of anisotropic pressure and always isotropise the pressure locally.} However, once the assumption of perfect spherical symmetry is relaxed, we expect the same physics as in the dust model is reproduced in a fluid with $w\ll 1$. That is, if one includes anisotropies in the perfect fluid model and performs coarse-graining to obtain an effective radial flow, the anisotropic pressure of this effective fluid recreates the effects observed in a dust model.}

\begin{center}
        
        \begin{figure}

    \begin{tikzpicture}[scale=1.8]
        \draw[darkgray,->,ultra thick] (0,0)--(7,0);
        \draw[darkgray,->,ultra thick] (0,0)--(0,5);
        \node[darkgray] at (3.5,-0.5) {$\ln a$};
        \node[darkgray] at (-0.5,3.5) {$\ln{r_{\rm p}}$};
        \filldraw[darkgray] (0,0) circle (0.02cm);
        
        \draw[thick,teal] (0.5,0)--(3.5,4.5);
        \node[teal,scale=0.8] at (2,3) {$H^{-1}\propto a^{3/2}$};
        
        \node[scale=0.8,align=left] at (3.6,2.3) {physical size of\\ perturbation};
        \draw[thick] (0,0)--(1.5,1.5);
        	\draw[dashed] (1.5,1.5)--(4,4);
        	\draw[thick] (1.5,1.5) arc (135:90:5);       	
        	\filldraw (5.03,2.96) circle (0.05cm);
        	\node[scale=1] at (5,3.2) {$r_{\rm max}$};
        	\draw[thick] (5.03,2.96) arc (90:30:1);
        	\filldraw (5.89,2.46) circle (0.05cm);
        	\node[scale=1] at (6,2.3) {$r_{\rm min}$};
        	
        	\draw[thick] (5.89,2.46)--(6,2.7);
        	\draw[thick] (6,2.7)--(6.5,2.7);
        	\node[scale=1] at (6.8,2.8) {$\dfrac{r_{\rm max}}{2}$};
        	
        	\draw[densely dotted] (5.03,2.96)--(5.03,0);
        	\node[scale=0.8] at (5.03,-0.3) { $a(t_{\rm max})=\dfrac{1}{\delta_m}\dfrac{3(6\pi)^{2/3}}{20}\approx \dfrac{1.06}{\delta_m}$};
        	
        	\draw[densely dotted] (1.5,1.5)--(1.5,0);
        	\node[scale=0.8] at (1.5,-0.2) {$a_{en}=1$};
        	
        	\draw[densely dotted,thick,purple] (0,1.5)--(6,1.5);
        	\node[purple,scale=0.8] at (-1,1.5) {$r_{\rm sh}=2GM$};
    \end{tikzpicture} 
            \caption{Evolution of a perturbation in MD era, $r_{\rm sh}$ is Schwarzschild radius, $r_{\rm min}$ is the raduis where the collapse stops, if $r_{\rm min}$ becomes smaller than $r_{\rm sh}$ then a black hole is formed, otherwise it will re-expand and reaches to $r_{\rm max}/2$ which is the equilibrium radius predicted by Virial Theorem.}\label{Fig1:FlcEvl}
        \end{figure}
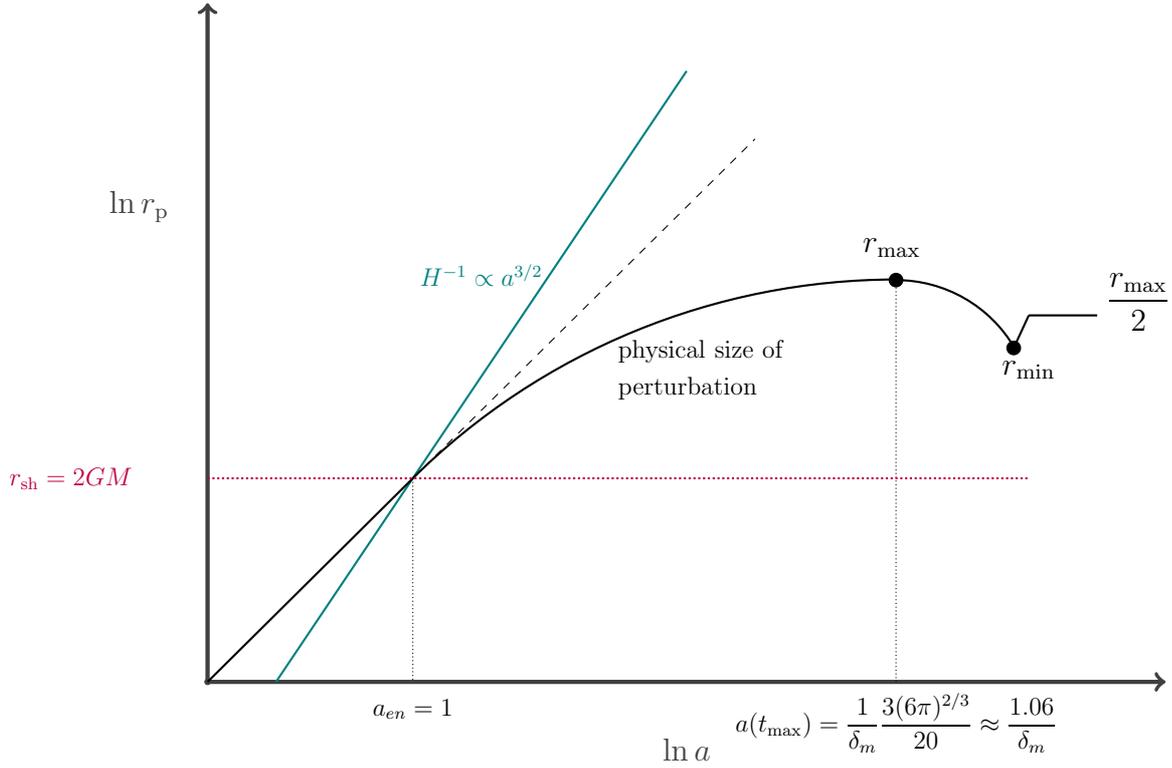
\end{center}
\section{Role of the perturbations}\label{sec:Pertb}
\subsection{Top-hat background}\label{sec:tophat}
As a warmup, we first study the role of perturbations if superposed on a ``top-hat'' density distribution, a homogeneous over-density within a radius $R(t)=R(r_m,t)$. Later, we will generalize the argument to other profiles. The evolution of the top-hat is exactly solvable \cite{Gunn:1972sv}, fully specified by its amplitude, $\delta_m$, and mass, $M$, or size, $r_m$. Based on the definitions of $\delta_m$ and $r_m$ in Section \ref{sec:RV}, the equivalent curvature profile is:
\begin{equation}
    K(r)=\left\{\begin{array}{lr}
        \frac53\frac{\delta_m}{r_m^2} & r<r_m\\
        \frac53\frac{\delta_m r_m}{r^3} & r>r_m 
    \end{array}
    \right. .
\end{equation}
Fig. \ref{Fig1:FlcEvl} shows the evolution of such a perturbation. Early on, the radius of the sphere increases with the scale factor of the universe. We assign to it a comoving wave-number
\be
k_0 := \lim_{t\to 0}  \frac{a(t)}{R(t)}=\frac{1}{r_m},
\ee 
and choose $a(t)= (3 k_0 t/2)^{2/3}$, which leads to $H_0 r_m=1$. With this normalization, the perturbation enters the horizon at $a=1$. The expansion of the sphere will eventually stop and its radius will reach
\be
R_{\rm max} = \frac{3 r_m }{5 \delta_m}.
\ee
This happens at the time $t_{\rm max}$ when
\be\label{amax}
a(t_{\rm max}) \approx \frac{1.06}{\delta_m} .
\ee
The sphere begins to collapse. A uniform sphere will collapse completely to $R=0$ and become a black hole, regardless of how small \(\delta_m\) is. It is useful to define a second scale factor for the contracting phase
\be\label{eq:bdef}
b(t) := \frac{R(t)}{R_{\rm max}}.
\ee
If $\delta_m \ll 1$, there is a long period of contraction before black hole formation at $R(t) = 2 GM=r_m$:
\be\label{bBH}
b_{\rm BH} = \frac53 \delta_m \ll 1.
\ee
Realistically, there are additional perturbations, represented by $\zeta_{\rm rms}$, that overlay this spherical perturbation. During the evolution of the top-hat background, fluctuations with wavelengths shorter than its size grow and induce random velocities, a velocity dispersion $\sigma_v$. If this grows large enough, then the centrifugal force can resist gravity:
\begin{equation}\label{eq:HaltCond}
    \frac{\sigma^2_v}{R(t)}\sim \frac{GM}{R(t)^2}.
\end{equation}
If this equality is reached while $R(t)>2GM$, the collapse will cease, and the sphere will begin to re-expand per the Virial theorem. Otherwise, a black hole forms. The smaller $\zeta_{\rm rms}$, the longer the inhomogeneities have to grow before \eqref{eq:HaltCond} is satisfied. Hence, the PBH formation threshold $\delta_{\rm th}$ will be smaller. In the following, we will find the parametric relation of $\delta_{\rm th}$ on $\zeta_{\rm rms}$, ignoring coefficients of $\mathcal{O}(1)$. 

The interior of the spherically symmetric top-hat solution is identical to a closed FLRW cosmology. One can directly solve for the evolution of perturbations within this cosmology (see Appendix \ref{sec:SPT}). The role of spatial curvature is important only near the turnaround time. Well before and well after, we can approximate it by flat FLRW cosmologies in the expanding and contracting phases. A typical perturbation with $k>k_0$, enters the horizon at $a_k = k_0^2/k^2$ and grows as 
\be
\delta_k(t)  \sim \zeta_k \frac{k^2}{k_0^2} a(t),\qquad t<t_{\rm max}.
\ee
We recall that in a matter-dominated universe, the equations of motion predict two solutions for matter perturbations: $ \delta_k\propto a(t) $, which represents a growing mode during an expanding phase, and $\delta_k\propto a^{-3/2}(t) $, which describes a decaying mode in that same phase. However, these solutions switch in a contracting phase and we have
\be\label{dkt}
\delta_k(t) \sim \delta_k(t_{\rm max}) b(t)^{-3/2} \sim \zeta_k \frac{k^2}{k_0^2} a(t_{\rm max}) b(t)^{-3/2},\qquad t>t_{\rm max}.
\ee
Modes with wave-number $k$ become nonlinear at the time $t_k$, when $\delta_k(t_k) = 1$. The velocity dispersion induced at that moment can be found using the Virial theorem. They have an enclosed mass $M_k \sim M (k_0/k)^3$ and size $R_k \sim R(t) k_0/k$:
\be\label{vk}
\sigma_{v,k}^2(t_k) \sim \frac{G M_k}{R_k}\sim \frac{G M}{R(t)} \frac{k_0^2}{k^2}.
\ee
Hence the condition \eqref{eq:HaltCond} is satisfied at $t_{k_0}$, when $\delta_{k_0}(t_{k_0}) \sim 1$. (We emphasize that $\delta_{k_0}$ and $\zeta_{k_0}$ denote typical perturbations at scale $k_0$, which are to be superposed on a much larger spherically symmetric peak.) Setting $k=k_0$ in \eqref{dkt} and requiring it to be $\mathcal{O}(1)$ at $b_{\rm BH}$ given by \eqref{bBH} and using \eqref{amax} results in the threshold 
\be\label{thr}
\delta_{\rm th} \sim \zeta_{k_0}^{2/5},
\ee
as originally derived in \cite{Harada:2022xjp}. {In Sections \ref{sec:Poisson} and \ref{sec:grid} we carry out numerical simulations to confirm this halting mechanism and to show how one can gain control over $b_{\rm min}$ by changing the amplitude of the perturbations in a top-hat overdensity. Note that these simulations are Newtonian, and do not determine the relativistic PBH-formation threshold. But they clearly show how $b_{\rm min}$ can change by varying $\zeta_k$.}

A potential caveat to the above argument is that we have neglected the blue-shifted velocity dispersion induced by the collapse of modes with $k>k_0$. Blueshifting \eqref{vk}, by a factor of $b(t_k)^2/b(t_{k_0})^2$ gives
\be\label{vk/vk0}
\frac{\sigma_{v,k}(t_{k_0})}{\sigma_{v,k_0}(t_{k_0}) } \sim \left(\frac{\zeta_k k_0}{\zeta_{k_0} k}\right)^{2/3},
\ee
which is less than $1$, except for a very blue spectrum $d(\log(k^3 P_\zeta(k)))/d\log(k)|_{k_0} \geq 2$. Since PBHs are more likely to form when $k^3P_\zeta(k)$ is maximum, it is safe to assume \eqref{vk/vk0} is less than $1$.

\subsection{General background}
\label{subsec:GB}
In the above derivation, we used the top-hat model to study the role of perturbations. {It would be helpful to summarize the arguments for the top-hat profile as follows. For a top-hat profile we have two scales: one is the size of the perturbation, $r_m$, the other is the nonlinear scale, $r_{\rm nl}^2\sim r_m^2 \zeta_{\rm rms}/(\delta_m b^{3/2})$, here we used Eq.\eqref{dkt} and $r_{\rm n.l}\sim k^{-1}$ when $\delta_k\sim 1$. The collapse will halt when the nonlinear scale reaches to the $r_m$, because when it happens condition \eqref{eq:HaltCond} holds. Requiring to form an apparent horizon at that moment leads to Eq.\eqref{thr}.} However, a typical profile is not flat, {this will introduce two other scales both within $r_m$, so can use the Taylor expansion of $K(r)$}. That is, we consider
\begin{equation}\label{eq:K}
    K(r)\approx \frac{\delta_m}{r_m^2}\Big(1-A_2\frac{r^2}{r_m^2}-A_4\frac{r^4}{r_m^4}+\cdots\Big).
\end{equation}
The reason why we explicitly include the term $r^4$ is that, as we saw in Section \ref{sec:wsol}, even without perturbations the PBH threshold is $\O(1)$ if $A_2=\O(1)$. Hence, we are interested in atypical peaks which are still much more likely than a top-hat, namely peaks with $A_2\ll 1$ but $A_4=\O(1)$. This introduces a new scale:
\begin{equation}
    r_{2-4}=r_m\sqrt{A_2/A_4} \ll r_m,
\end{equation}
above which the $r^4$ term dominates. Note that $r^3$ cannot contribute to the isotropic part of the profile, and the anisotropic contribution proportional to $r^3$ is $\mathcal{O}(\zeta_{\rm rms})$, which we regard as perturbations (see Appendix \ref{sec:sphr}).

\begin{figure}[t]
    \centering
    \includegraphics[width=0.9\linewidth]{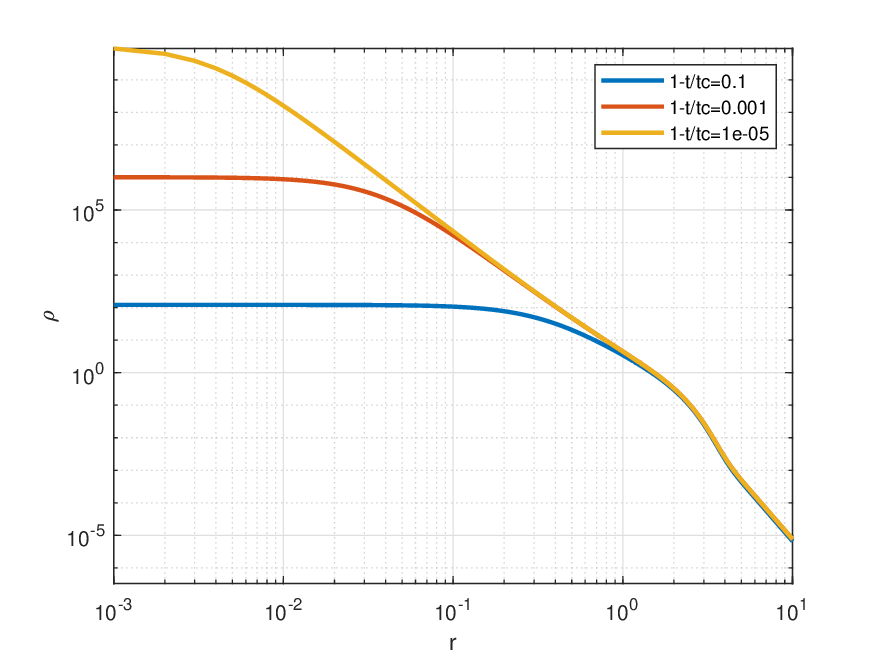}
    \caption{Three snapshots in the collapse of a Gaussian profile. The comoving radius of the central flat region $r_{\rm ed}$ shrinks from $r_m$ at $t_{\rm max}$ to $0$ at $t_c$.}
    \label{fig:edge}
\end{figure}

{The fourth scale is related the central flat region. For example, one can approximate the region $r\lesssim 0.3 r_m$ of a gaussian profile with a top-hat profile. We call this scale the ``edge'' scale, $r_{\rm ed}$. For the top-hat profile $r_{\rm ed}=r_m$ at all times. For a general profile, $r_{\rm ed}$ starts somewhere around $r_m$ and moves towards $r=0$ as we get closer to $t_c$ (defined in Eq.\eqref{eq:t1equation}) as in Fig. \ref{fig:edge}. We can find the evolution of $r_{\rm ed}$ by using Eq.\eqref{eq:sphsol1}. If we define $\delta t:=t_c-t$ and $\delta\theta:=2\pi-\theta$, for small $\delta\theta$ (i.e. deep in the collapsing phase) we can write }
\begin{equation}\label{eq:deltatheta}
    \frac{\delta\theta^3}{12\pi}\approx \frac{\delta t}{t_c}+\frac{3A_2}{2}\Big(\frac{r}{r_m}\Big)^2+\frac{3A_4}{2}\Big(\frac{r}{r_m}\Big)^4.
\end{equation}
{The central flat region is where the $r$-dependence in the above equation is negligible, that is $\frac{\delta t}{t_c}\gtrsim \frac{3A_2}{2}\Big(\frac{r}{r_m}\Big)^2+\frac{3A_4}{2}\Big(\frac{r}{r_m}\Big)^4$. This region can be approximated as a closed FLRW cosmology just like a top-hat profile. Inside this region, we have $b^{3/2}\approx 3\pi \delta t/2t_c$, and the location of the edge $r_{\rm ed}$ obeys the following relation}
\begin{equation}\label{eq:brrelation}
    b^{3/2}\sim A_2\Big(\frac{r_{\rm ed}}{r_m}\Big)^2+A_4\Big(\frac{r_{\rm ed}}{r_m}\Big)^4.
\end{equation}
Note that as we approach $t_c$, we have $b\to 0$ and the edge of the flat region reaches the center, {or $r_{\rm ed}\to 0$}.

However, before this occurs, perturbations inside the flat region grow, and similar to the top-hat case, when they become nonlinear the collapse will halt. {In other words, as $r_{\rm ed}\to 0$, we have also a growing $r_{\rm nl}$. The collapse will continue as long as $r_{\rm nl}<r_{\rm ed}$, but this condition will be violated at some point. When these two scales meet each other, the central region cannot collapse further due to velocity dispersion. Since the matter is non-interacting, if no black hole has formed at this moment, the infalling outer layers just passes through the central region without any interruption and re-expand. We expect no black hole formation after this point, as supported by the simulations discussed in Section \ref{sec:SimShell}.}

Hence, it suffices to check if an apparent horizon has formed at the time when the edge of the flat region coincides with the non-linear scale, i.e. the time $r_{\rm ed}=r_{\rm nl}$. To find it, we first use Eq.\eqref{dkt} and set $\delta_{k_{\rm nl}}\sim 1$  to find the non-linear scale $r_{\rm nl}:=1/k_{\rm nl}$. It satisfies
\begin{equation}\label{eq:nl}
    \frac{\zeta_{\rm rms}}{\delta_m}\sim A_2\Big(\frac{r_{\rm nl}}{r_m}\Big)^4+A_4\Big(\frac{r_{\rm nl} }{r_m}\Big)^6,
\end{equation}
where to simplify the equations, we assumed that the spectrum is sufficiently smooth at the scales of interest to replace $\zeta_{1/r_{\rm nl}} \to \zeta_{\rm rms}$. A posteriori, this will be justified by the fact that in practice, we will not have a huge hierarchy of scales. Then we set $r_{\rm ed}=r_{\rm nl}$ in Eq.\eqref{eq:brrelation} to find {the minimum compression ratio of the central region, i.e., $b_{\rm min}$}.

A black hole forms if at some radius $2GM/R$ reaches $1$. This ratio grows with $r$ as long as density is approximately uniform, i.e. $r<r_{\rm ed}$. For $r>r_{\rm ed}$, one can approximate $2GM/R$ using Eq. \eqref{eq:deltatheta}, Eq. \eqref{eq:brrelation}, and Eq.\eqref{eq:sphsol1}:
\begin{equation}\label{eq:phi}
    \frac{2GM}{R}=\frac{2K(r)r^2}{1-\cos\theta}\sim \frac{\delta_m(r/r_m)^2}{\Big[A_2\, (r/r_m)^2+A_4\, (r/r_m)^4\Big]^{2/3}}.
\end{equation}
Comparing $r_{\rm ed}=r_{\rm nl}$ and $r_{2-4}$ is important here. { It is because the behavior of $GM/R$ with $r$ is different for $r<r_{2-4}$ and $r>r_{2-4}$. For $r<r_{2-4}$ we have $GM/R\propto (r/r_m)^{2/3}$, so it will increase as we go to outer regions. But for $r>r_{2-4}$ the term with $r^4$ in the denominator dominates and we would have $GM/R\sim \delta_m (r/r_m)^{-2/3}$, so it decreases with $r$. Hence, if $r_{2-4} > r_{\rm ed}$ the maximum of $GM/R$ is around $r_{2-4}$.} Setting $A_4=1$ {for simplicity} leads to the following condition for black hole formation:
\begin{equation}\label{eq:THA2}
    \delta_m \gtrsim A_2^{1/3},\qquad r_{2-4}> r_{\rm ed}.
\end{equation}
It is important to note that this condition is not directly related to $\zeta_{\rm rms}$, but is indirectly related through the $r_{2-4}>r_{\rm ed}=r_{\rm nl}$ condition. On the other hand, if $r_{2-4} < r_{\rm ed}$ the maximum of $GM/R$ happens at the edge and Eq.\eqref{eq:nl} implies that the PBH formation condition is
\begin{equation}\label{eq:THzeta}
    \delta_m\gtrsim \zeta_{\rm rms}^{1/10}, \qquad r_{2-4}<r_{\rm ed}.
\end{equation}
Using this threshold, the condition $r_{2-4}<r_{\rm nl}$ translates into $A_2<\zeta_{\rm rms}^{3/10}$. Putting these together, with a profile of the form \eqref{eq:K} and $A_4 = \O(1)$, the threshold for PBH formation is 
\begin{equation}\label{eq:finalTh}
\delta_{\rm th}=\left\{ \begin{array}{cc}
  A_2^{1/3},& \qquad\zeta_{\rm rms} ^{3/10}<A_2<1 \\
\zeta_{\rm rms}^{1/10},& \qquad A_2<\zeta_{\rm rms}^{3/10}
\end{array}
\right..
\end{equation}
Note that for a parametrically small $\zeta_{\rm rms}$ the scales of interest are all $\ll r_m$ and hence the use of Taylor expansion is \eqref{eq:K} is justified. We can repeat the above analysis for an atypical profile $K(r)$, in which all $\{A_2,\cdots, A_{2n-2}\}\to 0$ while $A_{2n} = \O(1)$ to obtain
\be\label{eq:thn}
\delta_{\rm th}\sim \zeta_{\rm rms}^{~~~(4n-6)/10 n},
\ee
which in the $n\to \infty$ limit reproduces the top-hat result \eqref{thr}. However, as we will see in Section \ref{sec:Abund}, for finite $\zeta_{\rm rms}$ the power-law suppression from setting $\{A_2,\cdots,A_{2n-2}\}\to 0$ eventually wins over the exponential enhancement and the effective threshold will correspond to a finite $n$ in this equation. 

\subsection{Connecting $w\to0$ to $w=0$ in presence of perturbations}\label{sec:w0wneq0}
As mentioned in the previous section, at finite $w$, a spherically symmetric overdensity that passes the threshold $\delta_m \sim w$ collapses into a black hole due to the Jeans instability. We expect that once perturbations $\zeta_{\rm rms}$ are included, they lead to deviation from spherical symmetry and an effective shell crossing for small enough $w$. In the following, we will give a tentative estimate of this turning point, although a more dedicated analysis is necessary to confirm this picture. 

\begin{figure}[t]
    \begin{tikzpicture}[scale=2]
        \draw[darkgray,->,ultra thick] (0,0)--(7,0);
        \draw[darkgray,->,ultra thick] (0,0)--(0,5);
        \node[darkgray] at (3.5,-0.5) {$w$};
        \node[darkgray] at (-0.5,3.5) {$\delta_{\rm th}$};
        \filldraw[darkgray] (0,0) circle (0.02cm);

        	\draw[thick] (1.2,1.05) arc (115:105:30);      	
        \draw [thick] plot [smooth, tension=1] coordinates { (1.2,1.05) (1.1,1.01) (1,1)};

   		\draw [dashed] plot [smooth, tension=1] coordinates { (1,1) (0.6,2) (0.3,4.5)};
        
        \draw [dashed] plot [smooth, tension=1] coordinates { (1,1) (0.6,1.5) (0.3,2.5)};

        \draw [dashed] plot [smooth, tension=1] coordinates { (1,1) (0.6,1) (0.3,1.1)};

   	   	\node[scale=1] at (3,1.5) { $\sim w$};

		\draw[densely dotted] (1,1)--(1,0);
		\node[scale=0.9] at (1,-0.3) { $\mathcal{O}(\zeta_{\rm rms}^{2/5})$};
        
    \end{tikzpicture} 
            \caption{Schematic threshold for PBH formation for a typical high-amplitude profile. {For $\zeta_{\rm rms}^{2/5}\ll w\ll 1$, the threshold is linearly proportional to $w$. However, for $w\lesssim \zeta_{\rm rms}^{2/5}$, the effect of anisotropies become important. As $w$ is decreased well below $\zeta_{\rm rms}^{2/5}$, black hole formation threshold is fully determined by the profile shape, whose distribution is in turn fixed in terms of $\zeta_{\rm rms}$. More numerical work is needed to quantitatively determine the threshold beyond its parametric dependence on $\zeta_{\rm rms}$.}}\label{fig:th}
\end{figure}
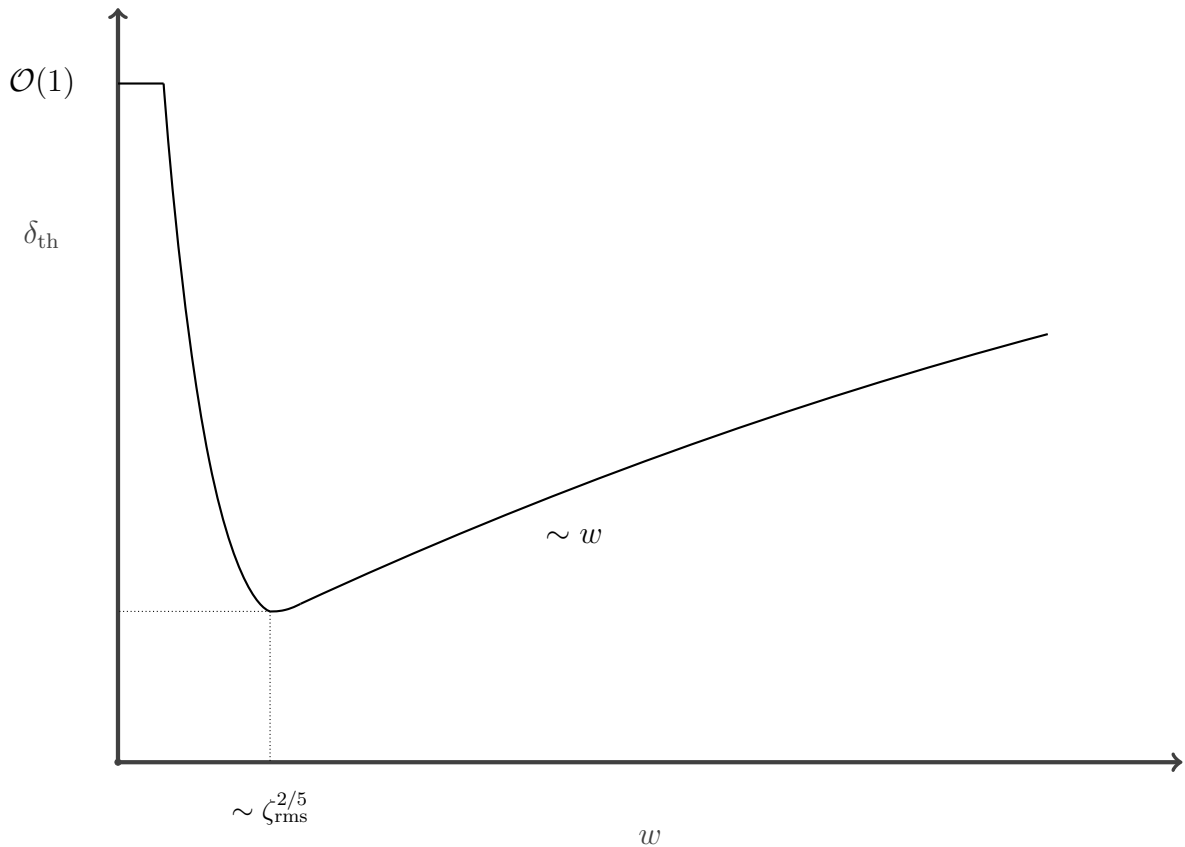

First, consider the evolution of perturbations before the turnaround time $t_{\rm max}$. \footnote{For a generic profile, the turnaround time is a function of $r$. Here we define $t_{\rm max}$ as the turnaround time at $r=r_m$. $r_m$ is 
defined above Eq.\eqref{eq:deltamdef}. } For $t<t_{\rm max}$, the comoving Jeans momentum, $k_J(t)\propto a(t)^{-1/2}$, reaches the minimum value
\be
k_J^0 := k_J(t_{\rm max}) \sim \frac{1}{r_m} \sqrt{\frac{\delta_m}{w}}.
\ee
{Note that $k_J^0$ is in fact the Jeans wavenumber at the turnaround time.} Perturbations do not grow below the Jeans length, hence the spectrum at $t_{\rm max}$ looks like
\begin{equation}
\delta_k(t_{\rm max}) =\left\{ \begin{array}{cc}
(k r_m)^2\dfrac{\zeta_{\rm rms}}{\delta_m}, & \qquad k<k_J^0\\
(k_J^0 r_m)^2\dfrac{\zeta_{\rm rms}}{\delta_m} \sim \dfrac{\zeta_{\rm rms}}{w}, & \qquad  k_J^0<k
\end{array}
\right..
\end{equation}
After $t_{\rm max}$, Jeans scale starts to shrink in the collapsing region (or grow in momentum space). This leads to a {power spectrum for smaller perturbations} that peaks at $k_J^0$, which for $\delta_m \sim w$ (the PBH threshold for a spherically symmetric overdensity) is essentially the total size of the structure: $k_J^0 r_m \sim 1$. We can determine if a black hole has formed by the time this scale becomes nonlinear. Of course, for a typical profile with $A_2= \O(1)$, the comoving scale $r_m$ will soon be far outside of the approximately flat central region. However, one might expect that perturbations still grow as $b(r,t)^{-3/2}$ where $b(r,t)= R(r,t)/R_{\rm max}(r)$ is the contraction of the shell at $r$ from its maximum physical radius (see appendix \ref{sec:SPTonsphere}). As such, the peak of the spectrum becomes nonlinear at $b_{\rm min} \sim \left({\zeta_{\rm rms}}/{w}\right)^{2/3}$, at which point
\be
\left.\frac{GM}{R}\right|_{r_m,t_{\rm nl}} \sim \frac{\delta_m}{b_{\rm min}} \sim\frac{w^{5/3}}{\zeta_{\rm rms}^{2/3}},
\ee
where we used $\delta_m\sim \delta_{\rm th} \sim w$ in the last expression. This estimate suggests that for $w<\zeta_{\rm rms}^{2/5}$, nonlinearity of perturbations plays a significant role and $\delta_{\rm th}$ should start to deviate from $\O(w)$. We expect that when $w \sim \zeta_{\rm rms}$, there will be no significant difference between dust and fluid. {For $w=\O(1)$, the dependence of PBH formation threshold on the profile shape has been studied in the literature (see, for example, \cite{Escriva:2019phb}). However, this dependence is more pronounced for softer equations of state; as we saw with the dust model, the threshold can change dramatically depending on the profile shape. Also, to extract the exact shape of the $\delta_{\rm th}(w)$ for $w\lesssim \zeta_{\rm rms}^{2/5}$, one needs to perform numerical simulations with certain assumptions about the shape of the profile. These ambiguities are reflected with the dashed-lines in Fig. \ref{fig:th}.}

\section{Numerical Confirmation of Halting Mechanism}\label{sec:NumAnl}
In this section, we will use Newtonian N-body simulations to provide evidence not for the threshold of PBH formation, which would require relativistic simulations like those in \cite{Lim}, but for the halting mechanism based on the growth of velocity dispersion. {We are going to carry out a series of simulations in which a spherical distribution of particles are going to collapse under Newtonian gravity.} In Sections \ref{sec:Poisson} and \ref{sec:grid}, we consider a top-hat background, where inhomogeneities are essential to prevent black hole formation. {Note that these top-hat distributions are not necessarily cosmological. They are neither expanding nor collapsing at the initial step, but soon due to gravity they will start to collapse. Due to finite particle effect, they naturally contain inhomogeneities which will have different power spectrum due to the different initial conditions imposed on particles position and velocities.} In Section \ref{sec:SimShell}, we consider {a collection of spherical shells, {initially at rest,} that follows a Gaussian density profile. {Then we compare the result with a collection of $N$-body simulations with the same density profile but different $N$s.} Here in Section \ref{sec:SimShell}, our aim is to show that even without small-scale anisotropies, spherical shell-crossing in a dust model will generically prevent PBH formation}.

In the first two simulations, we start with a sphere with $GM=1$ and $r=r_{\rm max}=1$. Then, we divide the mass between $N$ equal-mass particles and let it collapse until it starts to re-expand. We are going to compare the compression factor, $b_{\rm min}=r_{\rm min}/r_{\rm max}$ with the theoretical predictions. In the thi  rd case, we compare the 3+1 and 1+1 simulations. Also note that in these simulations, $t=0$ corresponds to $t=t_{\rm max}$ in the previous section.

\subsection{Poisson spectrum}\label{sec:Poisson}
In our first example, we distribute the total mass among the $N$ particles and allow it to collapse with zero initial speed. Setting the initial velocity to zero ensures that the power spectrum is completely determined by the initial positions. The initial positions of these particles are distributed with a uniform probability, so the power spectrum would be Poisson: $P_{\delta}(k)\sim \bar{n}^{-1}$ where $\bar{n}$ is the mean number density. Thus, the variance of perturbation on the size of the sphere is $\delta^2_{k_0}\sim N^{-1}$. In Section \ref{sec:tophat}, we discussed that when perturbations at scale $k_0$ become nonlinear, the collapse stops. Since this variance will grow with $b^{-3}$, then we predict the minimum contraction:
\begin{equation}\label{eq:halting1}
    b_{\rm min}\sim \frac{1}{N^{1/3}}.
\end{equation}
The Poisson spectrum of $\delta$ is equivalent to a primordial spectrum $P_\zeta(k) \propto k^{-4}$, which is red enough for this estimate to be justified. 

\begin{figure}[t]
    \centering
    \includegraphics[width=0.9\linewidth]{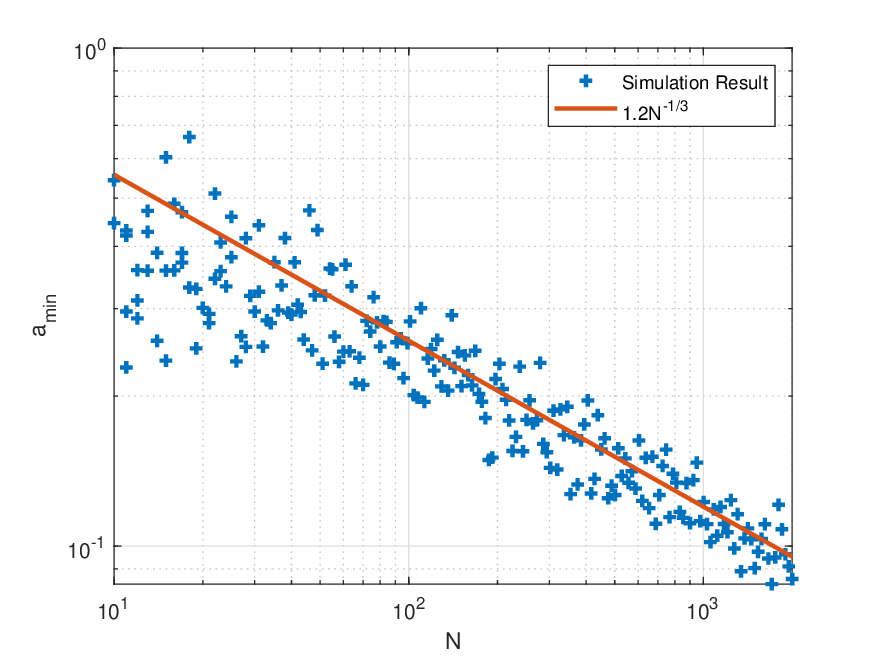}
    \caption{Each $+$ shows a $b_{\rm min}$ for a certain $N$ and the solid line is $1.2 N^{-\frac13}$.}
    \label{fig:Poisson}
\end{figure}

Fig.\ref{fig:Poisson} shows the result. Each point represents the compression factor of a simulation with certain $N$, and the solid line is $1.2\times N^{-1/3}$. Despite some scatter, Fig. \ref{fig:Poisson} shows that the compression factor agrees well with the estimation from \eqref{eq:halting1} across multiple orders of magnitude.

\subsection{Uniform sphere with the particles on cubic grid}\label{sec:grid}
For the next example, we set the initial position of $N$ particles on a regular cubic grid inside the sphere. We do this to minimize the initial density perturbations and gain control over the power spectrum through the velocity initial condition. For the initial velocity, we use a random uncorrelated Gaussian distribution for each particle with $\sigma_{v,i}(0)=s_v/\sqrt{3}$ so:
\begin{equation}
    v_k(0)\sim \frac{k^{3/2}}{\sqrt{\bar{n}}}s_v  \sim \Big(\frac{k}{
    {k_0}}\Big)^{3/2} \frac{s_v}{\sqrt{N}}.
\end{equation}
We expect that the displacement of particles creates new density perturbations. In particular, if
\be\label{N23}
s_v^2 > \frac{\delta_m}{N^{2/3}} = \frac{GM}{r_{\rm max} N^{2/3}},
\ee
After one dynamical time $t_d= (G M/r_{\rm max}^3)^{-1/2}$, the displacement would be larger than the lattice spacing and hence the mass distribution becomes essentially Poisson. 

However, it must be noted that due to the finiteness of the grid size on the scale of the sphere ${k_0}$, there will be a density fluctuation with the amplitude of $\delta^{\rm grid} _{k_0}\sim 1/N^{2/3}$. This is because there are $O(N^{2/3})$ particles near the surface region, and each one contributes to the total density as an independent random variable; thus the variance in mass is $\sigma^2(M)\sim m^2 N^{2/3}$, where $m$ is the mass of each particle. Given that the total mass is $Nm$, the density fluctuations will be $O(N^{-2/3})$. Due to the presence of this perturbation, after a dynamical time, the gravitational potential, which is sourced by $\delta^{\rm grid}$, changes the velocity at scale ${k_0}$:
\begin{equation}\label{eq:deltaV}
    v_{k_0}^2 \sim \frac{{{\delta_m}}}{N^{4/3}}+ \frac{s_v^2}{{N}}.
\end{equation}
When $s_v^2\gg {\delta_m}/N^{1/3}$, the main contribution to the velocity spectrum comes from initial velocities, $s_v$, so the subsequent evolution is just the cosmological blueshift of the initial value at every scale. The condition for the dominance of the initial velocities is stronger than \eqref{N23}. However, the gravitational effect coming from the resulting Poisson distribution is still negligible compared to the cosmological blueshift of the initial velocity. Compared with \eqref{eq:HaltCond}, we find a minimum radius proportional to $s_v^2$:
\begin{equation}\label{eq:halting3}
    b_{\rm min}\sim \frac{s_v^2}{{\delta_m}} \qquad \text{if $s_v^2\gg \frac{{\delta_m}}{N^{1/3}}$.}
\end{equation}
On the other hand, if $s_v^2 \ll {\delta_m}/N^{2/3}$ then according to \eqref{eq:deltaV} the evolution of velocity at later times will be dictated by the grid fluctuations, $\delta^{\rm grid}$, and this means that we only need to follow the evolution of $\delta^{\rm grid}_{k_0}$:
\begin{equation}\label{eq:gridlimit}
    \delta^{\rm grid}_{k_0}(t)\sim \frac{1}{N^{2/3}b^{3/2}}\implies b_{\min}\sim N^{-4/9}.
\end{equation}
We can extend this analysis to slightly larger values of $s_v$ that are not too low to become completely irrelevant and are not too high to overcome the gravity of matter perturbations, up to $s_v^2\sim \delta_m/N^{1/3}$. To estimate the evolution of these initial velocities, we let $\delta$ evolve. At the beginning of the evolution, while still $b\sim 1 $, the change in density field is due to the initial velocities. After a dynamical time $t_d\sim ({\delta_m}{k_0}^2)^{-1/2}$ one can use the continuity equation, $\dot{\delta}\sim k.v$, to find the new density fluctuations due to the initial velocities:\footnote{In the Appendix \ref{sec:SPT} we discuss how to set the boundary conditions of perturbations at the turnaround point $r=r_{\rm max}$. N-body simulations presented here are similar to that. Since $\dot{\delta}\propto k.v$, having initial velocity results in having  $\dot{\delta}\neq 0$ at $t=0$. }
\begin{equation}
    \delta^2_{k_0}\sim \delta_{\rm grid}^2+\Big(k_0 v_{k_0} t_d\Big)^2=\frac{1}{N^{4/3}}+ \frac{s_v^2}{N{\delta_m}}
\end{equation}
and after that this new power spectrum will evolve $\propto b^{-3}$, reaching $1$ at 
\begin{equation}\label{eq:halting2}
    b_{\min}\sim \Big( \frac{1}{N^{4/3}}+ \frac{s_v^2}{N{\delta_m}} \Big)^{1/3}, \qquad \text{if $s_v^2\ll \frac{{\delta_m}}{N^{1/3}}$.}
\end{equation}
In particular, when $s_v\to 0$, the above expression becomes identical with Eq.\eqref{eq:gridlimit}.

\begin{figure}
    \centering
    \subfigure[$N=123$ Particle]{\label{fig:Sv100}\includegraphics[scale=0.9]{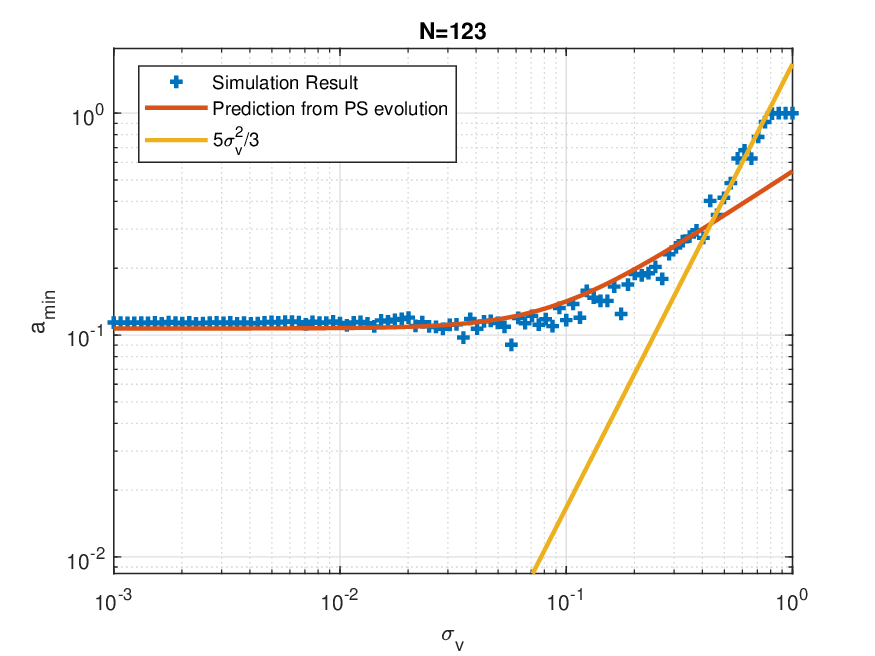}}
    \subfigure[$N=619$ Particle]{\label{fig:Sv600}\includegraphics[scale=0.9]{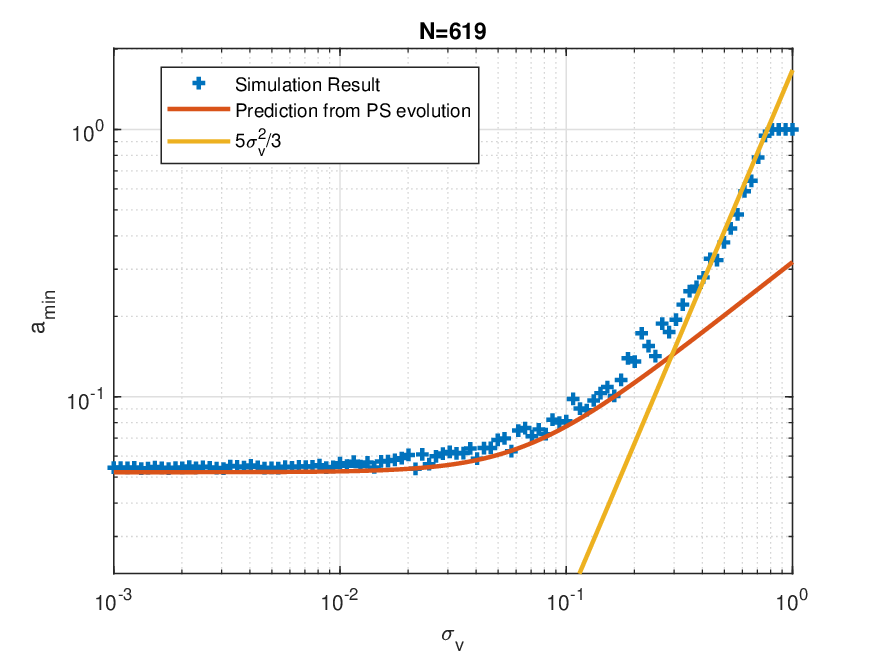}}
    \caption{Each point on the plots shows the results of simulations with a different $s_v$, the red curve represents Eq.\eqref{eq:halting2}, and the yellow line shows Eq.\eqref{eq:halting3}. }\label{fig:NvsSv}
\end{figure}

Fig.\ref{fig:NvsSv} shows the result of simulations for $b_{\min}$ compared to our estimations. The red curve shows the predicted value from Eq.\eqref{eq:halting2}, and the yellow line shows Eq.\eqref{eq:halting3}. Since we omitted numerical factors to get the red curve, we multiplied each term of \eqref{eq:halting2} with a certain coefficient of order unity, but for the different $N$, we used the same coefficients, which we believe are independent of $N$. The $\frac53$ coefficient of the yellow line comes from the Virial theorem, $\ddot{I}= 2K+U$ where $I=\frac12 \sum m \boldsymbol{r}^2$. In our setup $2K=M\sigma^2_v$ and, as shown in appendix \ref{sec:SPT}, $U=-\frac{3}{5}\frac{GM^2}{r_{\rm max}}$. Setting $GM =1= r_{\rm max}$, this means that at $\sigma^2_v=3/5$ there won't be any contraction, i.e. $b_{\rm min}=1$.

The results of these simulations validate the main point: velocity dispersion at the scale $k_0$ prevents the collapse of a large mass distribution.
\subsection{Shell-crossing of general profile}\label{sec:SimShell}
Here we perform two simulations to show the effect of shell-crossing. The first simulation is the same N-body simulation code of Section \ref{sec:Poisson}, but with only one difference: instead of uniformly distributing particles, we use a Gaussian distribution with zero mean and unit variance to set the Cartesian coordinates of each particle. Also, like \ref{sec:Poisson}, the initial velocity is set to zero. This is equivalent to a density profile proportional to $e^{-r^2/2}$ at rest. We create 100 realizations with different particle numbers, $N$, from $N=100$  to $N=2000$, and let them evolve under gravity and measure the minimum compression factor. The definition of the compression factor $b$ is the ratio of the mean radius of the system to its initial value.

The second simulation is a collection of spherical concentric shells with the same initial density profile as the first simulation. Each shell has a fixed mass and feels only the gravity of its inner shells. They evolve under Newtonian gravity with Plummer smoothing at $r=0$, and the time evolution is carried out by the leapfrog method. We use a fixed number of shells ($10000$ shells) to ensure accuracy. {Since it is the Plummer smoothing that softens the central divergence and allows the shells to cross the origin, $b_{\rm min}$ is expected to depend, in principle, on the smoothing length $\epsilon_{\rm p}$. In these units we set $\epsilon_{\rm p}=0.003$ we have checked that $b_{\rm min}\approx0.41$ is stable to within a $3$ percent as $\epsilon_{\rm sm}$ is varied over two orders of magnitude around $\epsilon_{\rm p}=0.003$, indicating that the result is not sensitive to the details of the softening, though we have not performed a full convergence study.}

\begin{figure}
    \centering
    \includegraphics[width=0.9\linewidth]{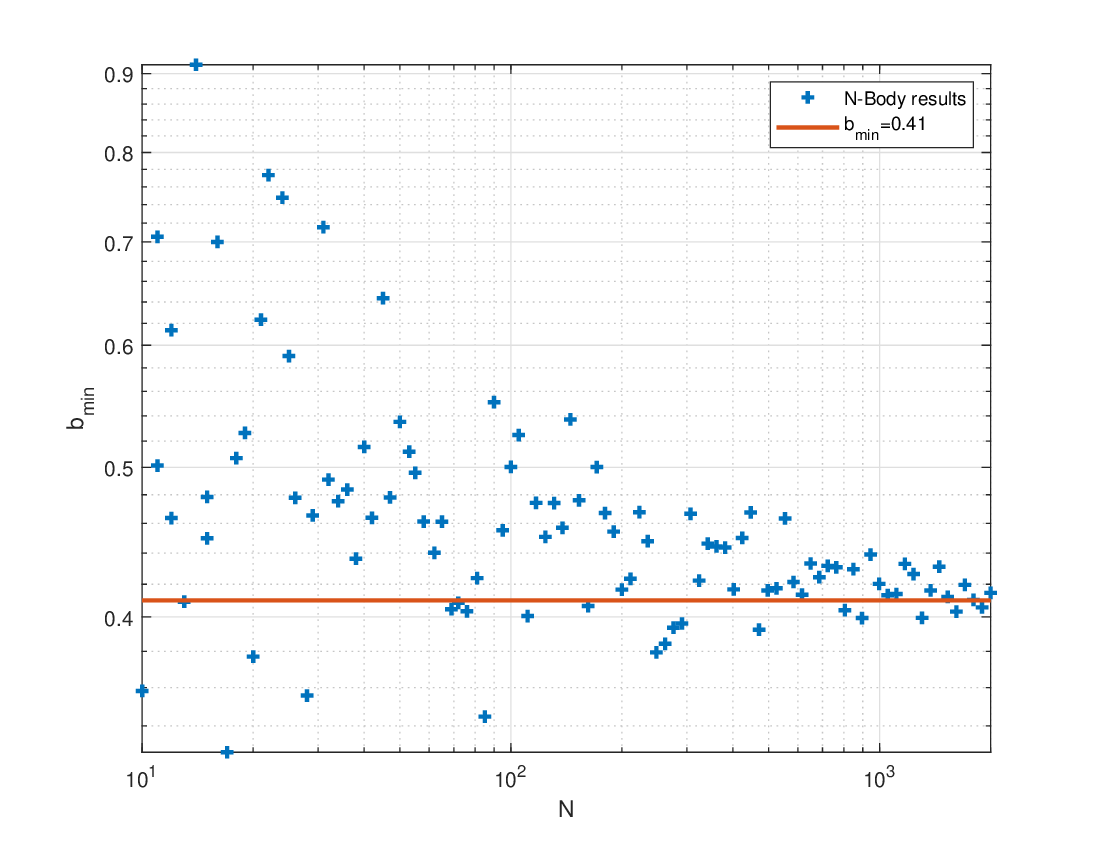}
    \caption{$b_{\rm min}$ for N-body simulations with Gaussian Density profile}
    \label{fig:bminGaussian}
\end{figure}

Fig.\ref{fig:bminGaussian} shows the results of these two simulations. Each point shows $b_{\rm min}$ for an $N$-particle realization, the solid line shows the value $b_{\rm min}$ from the shell simulation added for comparison. While for lower values of $N$ in the first simulations the $b_{\rm min}$ is scattered, for the larger $N$ the $b_{\rm min}$ shows a convergence to the value of second simulation $b_{\rm min}\approx0.41$. The tendency to higher values in small $N$ is due to the larger power spectrum of the higher Poissonian noise. Note that in the second simulation, we only have perfect spherical shells that can pass through each other. Thus, we can deduce that the minimum value of $b_{\rm min}\approx0.4$ in the N-body simulations comes from the shell-crossing. $b_{\rm min}$ generally depends on the shape of the profile, but for a typical profile with $A_2=\mathcal{O}(1)$, it is an $\mathcal{O}(1)$ number.
\section{Abundance and Clustering}\label{sec:Abund}
The potential significance of PBHs as both a source of gravitational wave background and a candidate for dark matter \cite{LISACosmologyWorkingGroup:2023njw} depends crucially on their population and mass distribution. We have argued that the black hole formation threshold for typical peaks is $\O(1)$, while exceptionally flat peaks can have lower thresholds. In particular, Eq. \eqref{eq:finalTh} shows the parametric behavior of the threshold assuming a suppression of the first term in the Taylor expansion of the initial overdensity. Below, we will use it to derive the parametric behavior of PBH energy fraction $\beta$, showing that for an observationally interesting value of $\beta$, PBHs formed from the collapse of such peaks dominate.

\begin{figure}
    \centering
    \includegraphics[width=0.9\linewidth]{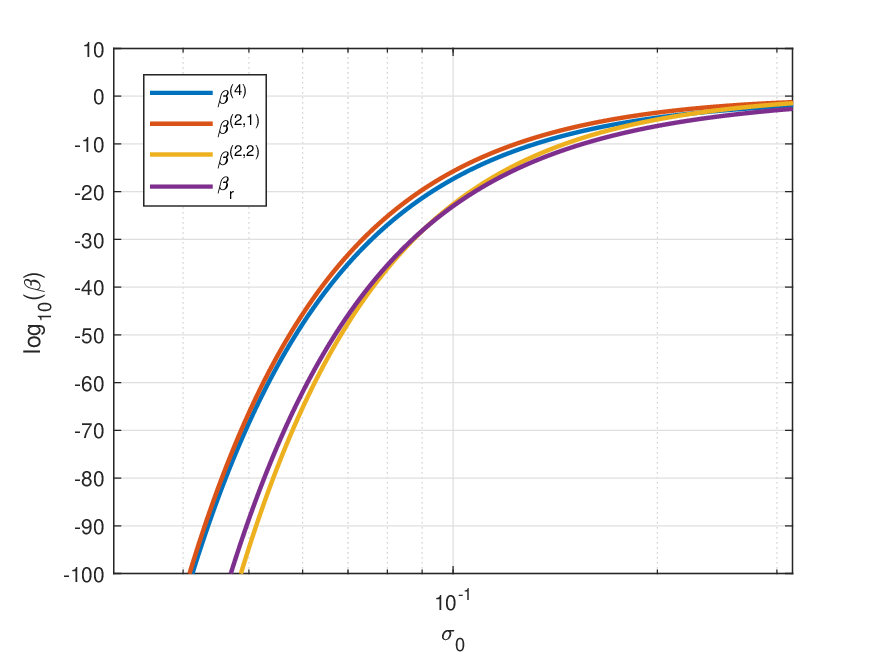}
    \caption{The numerical integration of PBH production probability in different regimes. $\beta^{(4)}$ is integration over $x<\sigma_0^{-3/5}$, $\beta^{(2,1)}$ is over $\sigma_0^{-3/5}<x<\sigma_0^{-1}$ and $\beta^{(2,2)}$ carried out over $\sigma_0^{-1}<x$. {Also we added $\beta_r$ to represent the parametric dependence of $\beta$ on the amplitude of a broad power spectrum in the radiation dominated case. To make it comparable with matter dominated case, we just set $\delta_c=1$ for radiation dominated, where $\delta_c$ is the threshold of PBH formation.}  }
    \label{fig:beta}
\end{figure}

{According to the peak theory \cite{BBKS}, the abundance of peaks above a certain threshold is exponentially sensitive to the threshold. Since we showed that flatter peaks can form a PBH with lower amplitudes, one must calculate the probability of the existence of a flat, low-amplitude peak capable of forming a black hole. We will assume a broad power spectrum.} First, we translate the conditions in the language of the peak theory (see Appendix \ref{sec:sphr}):
\begin{equation}
    x\sim A_2\nu,\quad \nu=\frac{\delta_m}{\sigma_0}, \quad \sigma_j^2=4\pi\int dk k^2 k^{2j} P_\zeta(k).
\end{equation}
Here we assumed $r_m^2\sigma_2/\sigma_0\sim 1$. By definition, $\sigma_0=\zeta_{\rm rms}$. {Smaller values of $x$ corresponds to flatter peaks.} The condition Eq.\eqref{eq:finalTh} in terms of these variables will be:
\begin{equation}
\nu_{\rm th}=\left\{ \begin{array}{cc}
  \sigma_0^{-1} &  \sigma_0^{-1}<x\\
  \Big(\frac{x}{\sigma_0^3}\Big)^{1/4}& \sigma_0^{-3/5}<x<\sigma_0^{-1}\\
\sigma_0^{-0.9}&   x<\sigma_0^{-3/5}
\end{array}
\right..
\end{equation}
Since the characteristic values of $x$ are  all  $\gg 1$, the joint probability distribution of the $\nu$ and $x$ among peaks will be approximately
\begin{equation}\label{eq:PeakDensity}
 n(\nu,x)   \propto x^3 e^{-\frac{\nu^2+x^2-2\nu\gamma x}{2(1-\gamma^2)}},\quad \gamma:=\frac{\sigma_1^2}{\sigma_0\sigma_2},
\end{equation}
with a dimensionful proportionality factor that cancels in computing $\beta$. We estimate the production probability by using a modified version of Press-Schechter formalism \cite{PS1974} to include the $x$ dependence of the threshold. The integration over $\nu>\nu_{\rm th}$ can be split into three integrals. One part comes from the integration of the region $x<\sigma_0^{-3/5}$. This is the region where the $A_4$ term dominates, so we call the contribution from this integral $\beta^{(4)}$. The $x>\sigma_0^{-3/5}$ is where the $A_2$ term determines the threshold, so we call $\beta^{(2,1)}$ the integration of $\sigma_0^{-3/5}<x<\sigma_0^{-1}$, and $\beta^{(2,2)}$ the contribution of $x>\sigma_0^{-1}$. {The total fraction {of energy stored in the form of PBH, $\beta:=(\rho_{\rm PBH}/\rho)_{\rm formation}$} is the sum of all above integrals: $\beta_{tot}=\beta^{(4)}+\beta^{(2,1)}+\beta^{(2,2)}$.} The asymptotic ($\sigma_0\to 0 $) behavior of the integrals can be estimated as
\begin{align}
    \beta^{(4)}&\sim \exp(- \kappa_4 \zeta_{\rm rms}^{-1.8}),\\
    \beta^{(2,1)}&\sim \zeta_{\rm rms}^{0.3}\exp (-\kappa_{2,1}\zeta_{\rm rms}^{-1.8}),\\
    \beta^{(2,2)}&\sim \zeta_{\rm rms}^{-2}\exp{(-\kappa_{2,2} \zeta_{\rm rms}^{-2})},
\end{align}
where $\kappa$'s are all $\O(1)$. The same approach applied to a radiation-dominated universe yields $\log\beta\sim -\zeta_{\rm rms}^{-2}$. Fig.\ref{fig:beta} shows the numerical integration of each $\beta$. {Note that Fig.\ref{fig:beta} is effectively assuming that all $\kappa$ values in the above equation are equal to 1.} If dark matter is composed entirely of PBHs with mass $m=10^{20}\rm{gr}$, {and the radiation dominated era starts right after the formation, which is the most conservative assumption about $\beta$,} then we must have  $\log_{10}\beta\sim-35$. This implies that $\zeta_{\rm rms}$ should be around $0.1$ for both the radiation-dominated and the matter-dominated cases. From Fig.\ref{fig:beta} it is evident that in $\log_{10}\beta\sim -35$ the $\beta^{(2,1)}$ is dominant and $\beta^{(4)}$ is also close. This is because most PBHs form due to the flatness of their profile. However, considering even flatter profiles as discussed around Eq. \eqref{eq:thn} is unnecessary because, despite weaker exponential suppression, those peaks are so rare that only for much smaller values of $\sigma_0$ and hence $\beta$ they start dominating. Hence, even though as $\sigma_0\to 0$ the PBH threshold asymptotes to $\zeta_{\rm rms}^{2/5}$, as derived in \cite{Harada:2022xjp}, for the realistic values of $\beta$, it is $\delta_{\rm th}\sim \zeta_{\rm rms}^{1/10}$. {Here we were interested in the parametric dependence of the abundance on the power spectrum amplitude and in identifying which term in Eq. \eqref{eq:finalTh} provides the dominant contribution to $\beta$. To compare the abundance of PBHs with the radiation-dominated case more quantitatively, one needs to determine the threshold more precisely, including the numerical coefficients that we have set to unity; this is most likely achievable via numerical simulations. }

Beyond the overall population, the clustering of PBHs can alter the formation rate of PBH binaries, consequently impacting the observable gravitational wave signatures\cite{Belotsky:2018wph,Chisholm:2011kn}. For PBHs formed during matter domination, our slightly lower-than-$1$ threshold implies some amount of subhorizon evolution during which the presence of a longer wavelength density perturbation can bias PBH formation just as in the case of halo formation. Note that in the absence of local non-Gaussianity, such an effect exists only if the long wavelength mode is also subhorizon at the time of the collapse. Hence, short subhorizon evolution before the collapse implies a small range of scales that can contribute to this bias. We will next estimate this bias, and then argue that it is completely inconsequential for PBH clustering.

The presence of a long mode $\delta_L$ changes the black hole formation threshold by modifying $\delta_{\rm th}$ in \eqref{eq:finalTh}. For any reasonable spectrum, this long mode will still be linear while the rare perturbation of size $1/k_0$ collapses. Hence, we can simply shift \eqref{eq:finalTh} with 
\be
\Delta\delta_{\rm th} \sim  \frac{\delta_L(t)}{a(t)}.
\ee
Reevaluating the above integral with the new threshold will give us a {\em positive} bias factor:
\begin{equation}
    \beta(t,x)=\bar{\beta}(1+b_1(t)\delta_L(t,x))\implies b_1(t)\sim \left(\frac{GM}{t}\right)^{2/3}\zeta_{\rm rms}^{-1.9}=\left(\frac{t_0}{t}\right)^{2/3}\zeta_{\rm rms}^{-1.9}
\end{equation}
where $t_0$ is the time at which $k_0= a(t_0) H(t_0)$. Hence, at $t_{\rm max}$ the bias is $\sim \delta_{\rm th}\zeta_{\rm rms}^{-1.9}$, which for $\zeta_{\rm rms}\sim 0.1$ gives $b_1(t_{\rm max}) = \mathcal{O}(10^2)$. This seemingly large bias results in an enhancement, compared to the Poisson distribution, of the probability of finding a second black hole within a Hubble patch given the presence of the first black hole (see e.g. \cite{Ali-Haimoud:2018dau,Desjacques:2018wuu}). However, it is of no consequence because the Poissonian prediction for this probability is $\sim 10^{-35}$, which remains negligible even after enhancement by a few orders of magnitude.

\section{Spin of Black Holes }\label{sec:Spin}
Having found the effective PBH formation threshold, we can estimate the RMS spin of PBHs formed in the matter-dominated era. The standard way to compute the spin of cosmological structures is to use the well-known Tidal Torque Theory \cite{1970Afz.....6..581D,White:1984uf}. According to TTT, tidal forces from surrounding matter perturbations exert torque on a proto-structure, causing its angular momentum to grow throughout the linear regime. This angular momentum reaches its maximum at turnaround. After that, the effect of these torques becomes negligible, and one can consider the angular momentum of the structure during the turnaround time as the final angular momentum of the structure.

The TTT starts with the Zeldovich approximation \cite{Zeldovich:1969sb}. Let $\boldsymbol{r}$ be the real position of particles and $\boldsymbol{q}$ be the Lagrangian coordinate of particles that are uniformly distributed. Then, the following relation is valid in linear approximation:
\begin{equation}\label{eq:Zeld1}
    \boldsymbol{r}=a(t)[\boldsymbol{q}-D_{+}(t)\mathbf{\nabla}_{\boldsymbol{q}}\phi],
\end{equation}
where $D_{+}(t)$ is the linear growth function and $\phi$ is the Newtonian potential. Here, the growth function of an expanding universe is relevant as we focus on the angular momentum of a sphere, which has been accumulated during its linear evolution prior to the turnaround time. Since the peculiar velocities of particles are known from Eq. \eqref{eq:Zeld1}, then one can write the total angular momentum by directly integrating $d L= \rho\,\boldsymbol{r}\times \boldsymbol{v}\, d^3\! \boldsymbol{r}$ over the Lagrangian volume $V_L$, and then subtracting the center of mass angular momentum. One would find:
\begin{equation}\label{eq:TTTI}
    L_i(t)=-a^2\Dot{D}_+\,\epsilon_{ijk} \frac{\partial^2\phi}{\partial q^l \partial q^k}\Bigg|_{q_{cm}} \int_{V_L} d^3\boldsymbol{q}\,\rho(t)a^3 \Delta q^j \Delta q^l \implies L_i= -a^2\Dot{D}_+\epsilon_{ijk} I_{jl}T_{lk} 
\end{equation}
$I_{ij}$ is the inertial tensor in the Lagrangian coordinate and $T_{ij}$ is the tidal tensor coming from surrounding matter perturbations. In a matter-dominated era, the time-dependent factor is equal to cosmic time $a^2\Dot{D}_+=t$. So one can estimate the final angular momentum knowing the turnaround time or $t_{\rm max}$: $ L_i\sim -t_{\rm max}\epsilon_{ijk} I_{jl}T_{lk} $. To find an estimate of spin, we compute the average $L^2$, using $\phi\sim \zeta$ in the linear regime:
\begin{equation}
    L^2\sim t_{\rm max}^2(3I_{ij}I_{ij}-(I_{ii})^2)\int_{k<k_0} d(\ln k) k^4 |\zeta_{\rm rms}|^2.
\end{equation}
Note that the momentum integral is effectively cutoff by the size of the region. One can rewrite the above expression:
\begin{equation}
L^2 \sim  GM^3 r_{\rm max} Q^2\sigma^2(M) ;\quad Q^2 :=\frac{3I_{ij}I_{ij}-(I_{ii})^2}{(Mr_{\max}^2)^2}
\end{equation}
where $Q$, which is proportional to the quadrupole moment of the structure, is a dimensionless parameter that vanishes for a perfect spherical perturbation. As mentioned earlier, rare peaks of a Gaussian distribution are nearly spherical. For an initial peak of height $\delta_m$, it can be shown that $Q\sim 1/\nu$, where $\nu \sim \delta_m/\zeta_{\rm rms}$ \cite{BBKS}. $\sigma(M)$ is $\delta_{\rm rms}(t_{\rm max})$ smoothed over a top-hat of radius $R=(3M/4\pi\bar{\rho})^{1/3}$ and one can show that $\sigma(M)\sim 1/\nu$. Therefore, we find
\begin{equation}
    L_{\rm rms}\sim \sqrt{GM^3 r_{\rm max}} \nu^{-2}\implies a_{\rm rms} =\frac{L_{\rm rms}}{2GM^2}\sim \frac{1}{\nu^2 |\delta_m|^{1/2}}
\sim \frac{\zeta_{\rm rms}^2}{{|\delta_m|}^{5/2}}
\end{equation}
where $a$ here is the dimensionless spin parameter of black holes. As discussed in Section \ref{sec:Abund}, most PBHs would form around $\delta_m\sim \zeta_{\rm rms}^{0.1}$. This leads to
\begin{equation}
    a_{\rm rms}\sim \zeta_{\rm rms}^{7/4}.
\end{equation}
Note that since in the radiation-dominated era the threshold is $|{\delta_m}|\sim 1$, the same estimation leads to $a_{\rm rms}\lesssim \zeta_{\rm rms}^2$, which agrees with \cite{Mirbabayi:2019uph}. For $\zeta_{\rm rms} \sim 0.1$, we have $a_{\rm rms} \sim 0.01$ in both MD and RD era.

\section{Conclusion}\label{sec:Con}
In this work, we investigated primordial black hole (PBH) formation during the matter-dominated era, incorporating key physical effects such as the dependence on the shape of the peak, shell-crossing, and velocity dispersion induced by small-scale perturbations. We argued that the PBH formation threshold for typical peaks is $\O(1)$ due to spherical shell-crossing, but exceptionally flat peaks can have lower thresholds that depend on the growth of anisotropies. Asymptotically it reaches $\delta_{\rm th}\sim \zeta_{\rm rms}^{2/5}$ for a top-hat perturbation. How flat the peaks that dominate PBHs formed in a given scenario are depends on $\zeta_{\rm rms}$ since there is a competition between flat profiles being rare at fixed amplitude and their lower threshold for black hole formation. We found that for cosmologically interesting densities of PBHs, the progenitors are peaks with exceptionally small second derivatives, but typical higher derivatives around the origin. To obtain reasonable densities of PBHs the scalar fluctuations have to be enhanced to $\zeta_{\rm rms} \sim 0.1$, which is several orders of magnitude larger than the CMB scales, and the effective PBH formation threshold is $\delta_{\rm th} \sim \zeta_{\rm rms}^{1/10}$. {Hence, in terms of abundance, for a broad power spectrum with the amplitudes typically used in the literature, there is little parametric difference between PBH formation during matter domination and radiation domination. More precise comparison between radiation and matter dominated cases needs performing numerical simulations.}

We also showed that contrary to earlier claims \cite{Saito:2024hlj,Harada:2017fjm}, the dimensionless spin parameter is small when $\zeta_{\rm rms}\ll 1$. If one holds the angular momentum responsible for preventing PBH formation, then it is reasonable to believe that most of the formed black holes must have large spins. However, the velocity dispersion effect is always larger than the angular momentum since the latter requires a net rotation. The spin of black holes can be detected in gravitational wave observatories \cite{Vitale:2014mka}, making it an interesting observational signature of the formation mechanism. 

Because of a period of sub-horizon growth before the collapse, PBHs are largely biased by long wavelength modes that are subhorizon at the time of the collapse. However, we argued that the effect of this on clustering is negligible. Finally, we mostly modeled non-relativistic matter by dust, and could only speculate about how including anisotropies will reconcile a fluid dynamic approach with small $w$ with our results. It would be interesting to further investigate this transition. Moreover, it would be interesting to further explore self-interactions, environmental and wave effects, which could be natural features of realistic early universe realizations of the scenario, on PBH formation and properties \cite{Flores,Easther,Deluca,Ralegankar}.

\section*{Acknowledgments}
We thank Yacine Ali-Haimoud, Paolo Creminelli, and Andrei Gruzinov for useful discussions. We would like to thank Ilia Musco for his detailed and insightful comments, which significantly improved the manuscript. We also thank the anonymous JCAP referee for their constructive feedback and questions, which helped us refine the paper.
\appendix

\section{Shape of the peaks}\label{sec:sphr}
Here we derive some statements about the shape of a high peak using methods of \cite{BBKS}. Assume a Gaussian random field and its Fourier transformation:
\begin{equation}
F(\boldsymbol{r})=\int d^3\boldsymbol{k} \,e^{i\boldsymbol{k}.\boldsymbol{r}}F(\boldsymbol{k}) ;\quad \braket{F(\boldsymbol{k}) F(\boldsymbol{k}')}=P(k)\,\delta^3_D(\boldsymbol{k}+\boldsymbol{k}').
\end{equation}
One can show that for every particular field configuration, the probability of that configuration, $\mathcal{P}$, is derived from the following action function:
\begin{equation}
\mathcal{P}[F]\propto e^{-S[F]}\implies S[F]=\frac{1}{2}\int d^3\!\boldsymbol{k} \frac{F(\boldsymbol{k})F(-\boldsymbol{k})}{P(k)}
\end{equation}
This is another manifestation that the field probability is Gaussian and homogeneous. Now, if we decompose the field in Fourier space into spherical harmonics:
\begin{equation}
F(\boldsymbol{k})=\sum_{l,m}C_{lm}(k)Y_{lm}(\hat{\boldsymbol{k}});\quad C_{lm}(k)=\int d\Omega F(k\hat{\boldsymbol{k}}) Y^*_{lm}(\hat{\boldsymbol{k}}),
\end{equation}
and the probability action will be:
\begin{equation}
S[F]=\frac{1}{2}\sum_{l,m} \int k^2 dk \frac{|C_{lm}(k)|^2}{P(k)},
\end{equation}
which means that $C_{lm}(k)$s are independent from each other. Using this, one can study the statistical properties of the shape of a peak. Also note that since $F(\boldsymbol{r})$ is real-valued we have: $C^*_{l,m}=(-1)^{l+m}C_{l,-m}$.

With the spherical-harmonic addition theorem:
\begin{equation}
e^{i\boldsymbol{k}.\boldsymbol{r}}=4\pi\sum_{m,l}i^l j_{l}(kr)Y_{ml}(\hat{\boldsymbol{r}})Y_{ml}^*(\hat{\boldsymbol{k}}),
\end{equation}
one can express the real-space value of $F$ in terms of $C_{lm}(k)$:
\begin{equation}
F(\boldsymbol{r})=\sum_{m,l} Y_{ml}(\hat{\boldsymbol{r}})\tilde{C}(r)_{lm};\quad \tilde{C}_{lm}(r)=4\pi i^l \int k^2 dk C_{lm}(k)j_l(kr).
\end{equation}
Now let's approximate the $F$ around $r=0$:
\begin{align}
    F(\boldsymbol{r})\approx & 4\pi Y_{00} \int dk k^2 C_{00}(k)+\frac{4\pi}{3} iY_{1m}r\int dk k^3 C_{1m}\\
    &-\frac{4\pi}{3} r^2 \Big(\frac{Y_{00}}{2}\int dk k^4 C_{00}(k)+\frac{Y_{2m}}{5}\int dk k^4 C_{2m}\Big)\\
    &+...,
\end{align}
where dots denote higher powers of $r$. To have a peak at $r=0$, one must set $\int k^3 C_{1m}(k)dk=0$, but also one needs to make sure that the last term is always negative. At this point,t it is better to write the above equation in Cartesian coordinates:
\begin{equation}\label{eq:Fapprox}
    F(\boldsymbol{r})\approx B+\eta_ix^i-\frac{1}{2}M_{ij}x^i x^j+...,
\end{equation}
where $\eta_i$ and $M_{ij}$ could be computed from integrals on $C_{1m}$ and $C_{2m}$ respectively. Having a peak is equivalent to $\eta_i=0$ and positive definite $M_{ij}$. One can write $M_{ij}$ in terms of spherical harmonics:
\begin{equation}
    M_{ij}=\frac13Q_{00}\delta_{ij}+\left(\begin{array}{ccc}
        Q_{2,2}+Q_{2,-2}-\sqrt{\frac23}Q_{20}&  (Q_{22}-Q_{2,-2})i & Q_{2,-1}-Q_{2,1}  \\
         & -Q_{2,2}-Q_{2,-2}-\sqrt{\frac23}Q_{20} & -(Q_{2,-1}+Q_{2,1})i \\
         & & 2\sqrt{\frac23}Q_{20}
    \end{array}\right)
\end{equation}
where:
\begin{align}
    Q_{00}=\sqrt{4\pi}\int dk k^4 C_{00},\quad Q_{2m}=\sqrt{\frac{2\pi}{15}}\int dk k^4C_{2m}.
\end{align}
Since we have $Q_{2m}^*=Q_{2,-m}(-1)^m$, then the $M_{ij}$ is real can be wrote in a simpler way:
\begin{equation}
    M_{ij}=\frac13Q_{00}\delta_{ij}+\left(\begin{array}{ccc}
        2\Re(Q_{22})-\sqrt{\frac23}Q_{20}&  -2\Im(Q_{22}) & -2\Re(Q_{2,1})  \\
         & -2\Re(Q_{22})-\sqrt{\frac23}Q_{20} & 2\Im(Q_{2,1}) \\
         & & 2\sqrt{\frac23}Q_{20}
    \end{array}\right).
\end{equation}
Only by looking at Eq.\eqref{eq:Fapprox} one sees 10 random variable: $B$, $Q_{00}$ and five $Q_{2m}$, and three $\eta_i$. Without imposing the positive definite condition, they are zero-mean Gaussian variables with the following parameters:
\begin{align}
    \braket{B^2}=4\pi \int dk k^2P(k)=\sigma_0^2, & & \braket{Q_{00}^2}=4\pi\int dk k^6P(k)=\sigma_2^2,\\ 
    \braket{BQ_{00}}=4\pi \int dk k^4 P(k)=\sigma_1^2,&  & \braket{Q_{2m}Q^{*}_{2m'}}=\frac{2\pi}{15}\int dk k^6 P(k)=\frac{\sigma_2^2}{30}\delta_{mm'}\\
      \braket{\eta_i\eta_j}=\frac{1}{3}\delta_{ij}\sigma_1^2 & & \sigma_j^2:=4\pi\int dk k^{2(1+j)}P(k).
\end{align}
Note that $Q_{20}$ is real so for $m\neq0$ we have $\braket{Q_{20}^2}=\braket{|Q_{2m}|^2}=2\braket{\Re(Q_{2m})^2}$. It is useful to define the following variables:
\begin{equation}
    \nu:=\frac{B}{\sigma_0},\quad x:=\frac{Q_{00}}{\sigma_2},\quad \gamma:=\frac{\sigma_1^2}{\sigma_0\sigma_2}.
\end{equation}
These are defined in a way that their variance is equal to $1$ and $\gamma$ is their correlation. Also definition of these variables is useful:
\begin{equation}
    y_0:=\frac{Q_{20}}{\sigma_2}, \quad y_{m}=\frac{\sqrt2\,\Re(Q_{2m})}{\sigma_2}\ \ \& \ \ y_{m+2}=\frac{\sqrt2\,\Im(Q_{2m})}{\sigma_2}\ \ \  {\rm for }\  m=1,2.
\end{equation}
Note that the variance of each $y$ is $1/30$, so the PDF of these variables is:
\begin{equation}
    f(\nu,x,y_m)=\mathcal{N} \exp-\frac12\Big(\frac{\nu^2+x^2-2\gamma x\nu}{1-\gamma^2}+30\sum_{j=0}^4y_j^2\Big),\quad \mathcal{N}=\frac{30^{5/2}}{\sqrt{(2\pi)^7(1-\gamma
    ^2)}}.
\end{equation}
Here we omit the distribution of $\eta_i$, but it is important to convert this PDF to a number density. Three $\eta_i$ are three independent Gaussian variables and to have a peak, means to be close to $\eta_i=0$ within the desired precision, $d^3\eta_i$. The probability of having $|\eta_i|<d\eta_i/2$ is:
\begin{equation}
    \frac{d^3\eta}{(2\pi/3)^{3/2}\sigma_1^3},
\end{equation}
the $d^3\eta_i$ can be interpreted as a volume of real-space that we are allowed to be there: $\eta_i$ around the peak can be approximated $\eta_i\approx M_{ij}x^j$ so changing the position means changing the $\eta_i$ but since $d^3\eta\approx |\det(M)|d^3x$ then if we move to a volume larger than $d^3\eta /|\det(M)|$ then $|\eta_i|<d\eta_i/2$ condition will be violated. So this is equivalent to having $\eta_{i}=0$ within volume $d^3x$, and multiplying this with the PDF for other variables will give us the number density. So the number density of points with $\eta_i=0$ is:
\begin{equation}
    n(\nu,x,y_m)dxd\nu d^5y= \bar{\mathcal{N}} |\det(M)| \exp-\frac12\Big(\frac{\nu^2+x^2-2\gamma x\nu}{1-\gamma^2}+30\sum_{j=0}^4y_j^2\Big)dxd\nu d^5y,
\end{equation}
where:
\begin{equation}
     \bar{\mathcal{N}}=\frac{3^{3/2} 30^{5/2}}{(2\pi)^5\sqrt{(1-\gamma
    ^2)}\sigma_1^3}.
\end{equation}
Now to convert the above expression to the number density of peaks, we should impose the positive-definiteness condition of $M$ and imposing this condition will change the Gaussian distribution. We know that a $3\times3$ matrix like $M$ is positive definite iff:
\begin{itemize}
    \item $Tr(M)>0$
    \item $Tr(M)^2>Tr(M^2)$
    \item $\det(M)>0$.
\end{itemize}
The first two conditions are equivalent to:
\begin{align}\label{eq:Cond1}
        Tr(M)>0\implies& x>0,\\ \label{eq:Cond2}
     Tr(M)^2>Tr(M^2)\implies &x^2>12\sum_{j=0}^4y_j^2,
\end{align}
which are easy to implement. However, $\det{M}>0$ is complicated. If we define the following variables:
\begin{equation}
    r_0=y_0,\quad r_1 e^{i\delta_m}=y_1+iy_3,\quad r_2e^{i\phi_2}=y_2+iy_4,
\end{equation}
the determinant can be written in this way:
\begin{equation}\label{eq:Cond3}
    \det(\frac{M_{ij}}{\sigma_2})=\Big(\frac{x}{3}\Big)^3-\frac{2x}{3}\sum_{j=0}^2r_j^2+\frac29\sqrt6r_0(2r_0^2+3r_1^2-6r_2^2)+2\sqrt2 r_1^2 r_2\cos(2\phi_1-\phi_2).
\end{equation}
As stated above, two of these conditions are easy to apply, but the last one related to $\det(M)$ is complicated. By using the Heaviside step function, $\theta$, one can integrate over $y_m$ and find the number density:
\begin{align}
      n(\nu,x)=\int d^5y\ \ n(\nu,x,y_m)\theta(x^2-12\sum y_m^2)\theta(g(x,y)),
\end{align}
where $g=\det(M)/\sigma_2^3$ (see \eqref{eq:Cond3}). We omit the $x>0$ condition since it plays no role in integrating over $y_m$, but it's a necessary condition to have a peak. General results for $n$ are complicated, but one can find its shape for certain limits. 

For $x\to \infty$ the positive definite condition is not important because it is unlikely for $y_j$ to reach values that violate the positive definiteness. So in this case, we have:
\begin{equation}
   n(\nu,x)\to \mathcal{N}_1 (x^3-3x) \exp-\frac12\Big(\frac{\nu^2+x^2-2\gamma x\nu}{1-\gamma^2}\Big),\quad \mathcal{N}_1=\frac{1}{3^{3/2}(2\pi)^{5/2}\sqrt{(1-\gamma
    ^2)}}\frac{\sigma_2^3}{\sigma_1^3}.
\end{equation}
But in the $x\to 0$ limit, the condition \eqref{eq:Cond2} forces $y_j< x/\sqrt{12}$. Then the integration over $\theta(g(x,y))$ should add a numerical factor independent of $x$, this leads to
\begin{equation}
   n(\nu,x)\sim \frac{ x^8\sigma_2^3}{\sigma_1^3} \exp-\frac12\Big(\frac{\nu^2+x^2-2\gamma x\nu}{1-\gamma^2}\Big).
\end{equation}
These results are in agreement with \cite{BBKS}. 

Now, we can investigate the sphericity of the peaks. To be more concrete, let us define the inertial tensor of the peak as:
\begin{equation}
I_{ij}=\int d^3\boldsymbol{r} r_i r_j F(\boldsymbol{r})
\end{equation}
then the spherical part is ${\rm Tr}(I_{ij})$:
\begin{equation}
I_{ii}=4\pi \int r^4 dr \tilde{C_{00}}(r)
\end{equation}
while the non-spherical part (that is relevant to the angular momentum calculations) is:
\begin{equation}
Q^2 I_{ii}^2=3(I_{ij})^2-(I_{ii})^2=\frac{8\pi}{5}\int dr dr' r^4 r'^4 \sum_{m=-2}^{2} \tilde{C}_{2m}(r)\tilde{C}_{2m}(r'),
\end{equation}
since the statistics of $C_{2m}$ is independent from $C_{00}$ and $C_{00}\propto \nu$, then one can deduce that $Q\sim 1/\nu$.

It is also possible to talk about the radial profile. The most probable radial profile can be calculated by minimizing $S[F]$ with the height constraint:
\begin{equation}
\frac{\delta}{\delta C_{0}} \Big(S[F]-\lambda (\nu\sigma-F(0))\Big)= \frac{k^2 C_0(k)}{P(k)}-\lambda \times 4\pi k^2=0,
\end{equation}
where $\lambda$ is the Lagrange multiplier of the height constraint and can be determined as follows:
\begin{equation}
C_0(k)=4\pi \lambda P(k) \implies \sigma \nu = (4\pi)^2 \lambda \sigma^2\implies \lambda =\frac{\nu}{\sigma (4\pi)^2}.
\end{equation}
This means that the maximum likelihood value of $C_{00}$ is:
\begin{equation}
C_{00}(k)=\frac{\nu}{4\pi \sigma } P(k)
\end{equation}
This means that the most probable profile in this case is actually the correlation function multiplied by a factor:
\begin{equation}
F(r)_{\max \mathcal{P}}=\frac{\nu}{\sigma}\xi(r);\quad \xi(r)=4\pi \int dk k^2 P(k) j_0(kr).
\end{equation}
There are perturbations in this profile of order $\sigma$, so the radius at which $\xi(r)\sim \sigma^2/\nu$ is the radius at which the approximations break down. The other way of seeing this is by looking at perturbations around $C_{00}$, let's define $\delta C_{00}$ as:
\begin{equation}
C_0(k)=\frac{\nu}{4\pi \sigma} P(k)+\delta C_0(k);\quad \int k^2 dk \delta C_0 (k)=0.
\end{equation}
Then the action for the probability of $\delta C_0$ is:
\begin{equation}
S[\delta C_0]=\frac{1}{2}\int k^2 dk \frac{|\delta C_0|^2}{P(k)}
\end{equation}
which is just typical fluctuations around some point (which is not necessarily around a peak), and this means a typical profile is correlation functions plus some fluctuations of order $\sigma$.

\section{Separate Universe Approach}
\label{sec:SPT}
The separate universe approach to perturbation theory suggests that a region of the universe, when smoothed over a suitable scale, evolves in the same way as the homogeneous background universe but with modified or renormalized parameters. This approach incorporates perturbations into the corresponding background parameters of an appropriate cosmological model, treating it as a separate universe with modified cosmology.

In particular, let us assume a flat fiducial cosmology evolves by the following equation
\begin{align}
\dfrac{\dot{a}^2(t)}{a^2(t)} =\dfrac{8 \pi G}{3} \bar{\rho}(t).
\end{align}
A smoothed overdense/underdense region of the universe with overdensity $\delta(t)$ evolves similarly to a closed/open universe. To the first order in $\delta$, one has \cite{Wagner:2014aka}
\begin{align}
\dfrac{\dot{\tilde{a}}^2(t)}{\tilde{a}^2(t)}+ \dfrac{\tilde{K}}{\tilde{a}^2(t)} =\dfrac{8 \pi G}{3} \tilde{\rho}(t)    
\end{align}
where
\begin{align}
&\tilde{\rho}(t)= \big(1+\delta(t)\big) \rho(t)
\\
&\tilde{a}(t)= \left(1-\dfrac{\delta(t)}{3}\right) {a}(t)
\end{align}
and effective curvature at some initial time $t_i$ can be found via
\begin{align}
-\Omega^i_k :=\dfrac{\tilde{K}}{\tilde{a}^2(t_i) \tilde{H}^2(t_i)} = \dfrac{5}{3}\,\delta_i =\dfrac{5}{3}\delta_m a(t_i)
\end{align}
We assume that at $\Omega_m(t_i) \simeq 1$. We denote the initial overdensity $\delta(t_i)$ simply as $\delta_i$ for brevity. Besides, to clarify, we used parameters with a bar, such as ${a}$ and ${\rho}$, to denote the parameters of the fiducial cosmology. In contrast, parameters with a hat are used to represent the "separate universe" parameters.

The above equation has the following parametric solution in terms of the conformal time $\theta$:
\begin{equation}
    \tilde{a}(t)=A(1-\cos\theta)\quad;t=B(\theta-\sin\theta)
\end{equation}
where:
\begin{equation}
    A= \frac{1-\Omega_k^i}{2|\Omega_k^i|}\quad; B=\sqrt{\frac{r_{\rm max}^3}{8GM}}=\frac{A}{\sqrt{|\Omega_k^i|}\tilde{H}_i}\quad; \tilde{H}_i=H_i\times(1-\frac13\delta_i)
\end{equation}
also note that in this parametrization, the value of $b$, defined by Eq.\eqref{eq:bdef} is determined by the following equation:
\begin{equation}
    b(t)=\frac{1}{2}(1-\cos\theta)
\end{equation}
Within this separate universe, the evolution of shorter perturbations is described by this equation:
\begin{equation}
    \frac{d}{d\theta}\Big((1-\cos\theta)\frac{d\delta}{d\theta}\Big)-3\delta=0.
\end{equation}
This equation has two independent solutions. If we set boundary conditions at $\theta=\pi$ or turn around time, we will have:
\begin{equation}\label{eqapx:generalsol}
    \delta(\theta)=2B\frac{d\delta}{dt}\Big|_{\theta=\pi}\times \frac{-4\cot\frac{\theta}{2}}{1-\cos\theta}+\frac{\delta|_{\theta=\pi}}{2}\times\left( \frac{6-3(\theta-\pi)\cot\frac{\theta}{2}}{1-\cos\theta}-1\right)
\end{equation}
in general, this solution is divergent at $\theta\to 0,2\pi$ which corresponds to $r\to 0$:
\begin{equation}
 \delta(0^+)\to \Big(\frac{3\pi}{4}\delta(\pi)-4B\dot{\delta}(\pi)\Big)b(t)^{-3/2};\quad \delta(2\pi^-)\to \Big(\frac{3\pi}{4}\delta(\pi)+4B\dot{\delta}(\pi)\Big)b(t)^{-3/2},
\end{equation}
so if we wish to connect this solution to a cosmological initial condition, we must set: 
\begin{equation}
    4B\frac{d\delta}{dt}\Big|_{\theta=\pi}=\frac{3\pi}{4}\delta|_{\theta=\pi}.
\end{equation}
This condition implies $\delta(\theta)=O(\theta^2)$, then if we normalize this solution to get the growth function, $D(t\to 0)= a(t)$, we will find that:
\begin{equation}
    D(\theta)=\frac{3}{2\delta_m}\Big(6\frac{1-\frac{\theta}{2}\cot{\frac{\theta}{2}}}{1-\cos{\theta}}-1\Big).
\end{equation}
The asymptotic behavior of this growth function in $t\to 2 t_{\rm max}$ or $\theta\to 2\pi$ is what we are interested in:
\begin{equation}
    D(\theta\to2\pi^-)=\frac{9}{\delta_m}\Big(\frac{\pi\sqrt{2}}{(1-\cos{\theta})^{3/2}}\Big)=\frac{9\pi}{2\delta_m}b^{-3/2}\approx \frac{36\pi}{\delta_m (2\pi-\theta)^3}
\end{equation}
Now we can confirm our estimation by these tools we have, first of all note that the Schwarzschild radius is equal to the size of the horizon, so if we set $t_i$ as the time of horizon entry, then $r_{ch}=H_i^{-1}$ and $\delta_i=\delta_m$. This means that once the mode of desire re-enters or crosses the horizon, its size is equivalent to the horizon size. We aim to examine the growth factor of short fluctuations from the moment when the long modes, $\delta_i$, enter the horizon until the corresponding mode exits the horizon before collapsing. At the time of horizon re-entry or crossing, we have the relationship $k = \tilde{a}(t) \tilde{H}(t)$. Therefore,
\begin{align}
\theta_{i}= \,\sqrt{\dfrac{20}{3}\delta_m};\qquad \theta_{c}=2\pi- \,\sqrt{\dfrac{20}{3}\delta_m}
\end{align}
These conformal time correspond to $t=t_i$ and $t=t_c := 2t_{\rm max}-t_i$ respectively. By which the growth factor between horizon re-entry and horizon ``re"-crossing was found to be by using the fact that $\theta_{i}\ll 1$ :
\begin{align}
    \dfrac{D(\theta_{c})}{D(\theta_{i})}\approx \frac{36\pi}{\delta_m \ (2\pi-\theta_c)^3}=\dfrac{27  \pi }{10 }\sqrt{\frac{3}{5}} ~{\delta_m}^{-5/2} \simeq \, 6.57~{\delta_m}^{-5/2}
\end{align}
Now, we can use this equation to find the corresponding condition of PBH formation at the time of horizon crossing. Using the Poisson equation and noting that $\tilde{\rho}(t)\sim \tilde{a}^{-3}(t)$, we find out that-- for sub-horizon fluctuations-- the amplitude of gravitational potential at horizon re-crossing is related to its value at horizon crossing via the following relation
\begin{align}
    \dfrac{\zeta_{\mathrm{rms}}(t_c)}{\zeta_{\mathrm{rms}}(t_i)}= \dfrac{\tilde{a}(t_i)}{\tilde{a}(t_c)} \dfrac{D(t_c)}{D(t_i)} = \dfrac{D(t_c)}{D(t_i)}.
\end{align}
In the last equality, we used the fact that $\tilde{a}(t_c) = \tilde{a}(t_i)$ and $\tilde{H}(t_c) = \tilde{H}(t_i)$. On the other hand, using the Friedmann equation, we have
\begin{align}
\dfrac{GM}{{r}(t_c)} = \dfrac{4\pi G}{3} {\rho}(t_c){r}^2(t_c)= \dfrac{1}{2}\tilde{H}^2(t_c) {r}^2(t_c) =\dfrac{1}{2}\dfrac{\tilde{a}^2(t_c)\tilde{H}^2(t_c)}{\tilde{a}^2(t_i)\tilde{H}^2(t_i)}=\dfrac{1}{2}  
\end{align}
It is already known that the gravitational potential on the black hole horizon is $\phi(R_{\mathrm{sch.}}) = -1/2$. Putting these all together, the condition for PBH formation, $\zeta_{\mathrm{rms}}(t_c)< \dfrac{GM}{\tilde{r}(t_c)}$, can be translated to those parameters at the horizon re-entry as
\begin{align}
    \zeta_{\mathrm{rms}}(t_i) \lesssim \dfrac{D(t_i)}{D(t_c)} \sim \delta_m^{5/2}
\end{align}
where in the last step we used the fact that $\frac{D(t_c)}{D(t_i)}\sim \delta_i^{-5/2}$. Therefore, the necessary condition for the overdense region to collapse to a PBH can be stated as
\begin{align}
     \delta_m > \zeta_{\mathrm{rms}}(t_i)^{2/5}
\end{align}
which is the result that we expected.

\section{Perturbation theory with spherical symmetry}\label{sec:SPTonsphere}
Here we wish to study the evolution of linear perturbations on a general spherical profile. One can add perturbations on top of a spherical matter over-density:
\begin{equation}
\rho(\boldsymbol{r},t)=\bar{\rho}(r,t)(1+\delta(\boldsymbol{r},t));\quad \boldsymbol{V}=U\hat{\boldsymbol{r}}+\boldsymbol{v};\quad U=\frac{\partial R}{\partial t},
\end{equation}
where $\bar{\rho}$ \& $U$ are background variables and $\delta$ \& $v$ are perturbations. Then the time evolution of these perturbations can be derived from the following equations:
\begin{align}
&\frac{\partial \delta}{\partial t}+\boldsymbol{\nabla .v}+v_r \frac{\bar{\rho}'}{\bar{\rho}R'}=0,\\
& \frac{\partial v_r}{\partial t}+\frac{U'}{R'}v_r=-\frac{\phi'}{R'},\\
&\frac{\partial\boldsymbol{v}_{\perp}}{\partial t	}+\frac{U}{R}\boldsymbol{v}_{\perp}=-\nabla_{\perp}\phi,
\end{align}
and the Background equations are:
\begin{equation}
\frac{\partial U}{\partial t}=-\frac{GM}{R^2};\quad \bar{\rho}R^2R'=\text{constant w.r.t}\ \ t.
\end{equation}
If $R$ and $R'$ follow a similar time evolution, then this is like standard perturbation theory in an isotropic background. To see this one can use the background solutions, Eq.\ref{eq:sphsol1}, to calculate $R'$:
\begin{align}
& t=\frac{H_0^2}{2K^{3/2}(r)}(\theta-\sin\theta) \implies \theta'=\frac{\theta-\sin\theta}{1-\cos\theta}\frac{3K'}{2K}\\
&\implies R'=R\Big(\frac{1}{r}-\frac{K'}{K}+\frac{3K'}{2K}\frac{(\theta-\sin\theta)\sin\theta}{(1-\cos\theta)^2}\Big).
\end{align}
It seems that the last term could diverge when $\delta\theta=2\pi-\theta\to 0$ and change the time evolution of $R'$. First note that this can't happen before $t_c$ because at this time $\delta\theta^3\sim \frac{K'r}{K}$ and this means that the last term could only be at the same order as the first term. To show that, let's define $\delta t=t_c-t$ then:
\begin{equation}
1-\frac{\delta t}{t_c}\approx \Big(\frac{K(0)}{K(r)}\Big)^{3/2}\Big(1-\frac{\delta\theta^3}{12\pi}\Big)\implies \frac{\delta\theta^3}{12\pi}\approx  \frac{\delta t}{t_c}+\frac{3A}{2}\Big(\frac{r}{r_m}\Big)^n,
\end{equation}
This shows that the difference between the time evolution of $R$ and $R'$ arises only when $\delta t$ is too small compared to $t_c$, and after that, shell-crossing will start.

\bibliography{PBHbib.bib}{}
\bibliographystyle{utphys}
\end{document}